\documentclass{article}
\usepackage{fancyvrb,fancyhdr}
\usepackage[english]{babel}
\usepackage{amsmath}
\usepackage{amsfonts}
\usepackage{amsthm}
\usepackage{graphicx}
\usepackage{mathtools}
\usepackage{csquotes}

\def\P{\mathbb{P}}
\def\E{\mathbb{E}}

\def\<{\langle}
\def\>{\rangle}

\theoremstyle{definition}
\newtheorem{exmp}{Model}

 \usepackage[style=authoryear,backend=biber]{biblatex}
\begin{document} 
\title{Stochastic Modeling and Statistical Inference of Intrinsic Noise in Gene Regulation System via the Chemical Master Equation}
\author{Chao Du $^{\ast }$, and Wing Hong Wong \thanks{%
Chao Du is Assistant Professor of Statistics, University of Virginia,
Charlottesville, VA 22904 (E-mail: cd2wb@virginia.edu). Wing Hong Wong is Professor of Statistics, Stanford University, Palo Alto, CA, 94305
(E-mail: whwong@stanford.edu)}}
\maketitle
\abstract{Intrinsic noise, the stochastic cell-to-cell fluctuations in mRNAs and proteins, has been observed and proved to play an important roles in cellular systems. Due to recent developments in single-cell-level measurement technology, the studies on intrinsic noise are becoming increasingly popular among scholars.  The chemical master equation (CME) has been used to model the evolution of complex chemical and biological systems since 1940, and is often put forth as the standard tool for modeling intrinsic noise in gene regulation systems. A CME-based model can capture the discrete, stochastic, and dynamical nature of gene regulation systems, and may offer causal and physical explanations of the observed data at single-cell level. Nonetheless, the complexity of the CME also poses a serious challenge for researchers in proposing practical modeling and inference frameworks. In this article, we will review the existing works on the modeling and inference of intrinsic noise in gene regulation systems within the framework of the CME model.  We will explore the principles of constructing a CME model for studying gene regulation systems and discuss the popular approximations of the CME. Then we will study the simulation simulation methods as well as the analytical and numerical approaches that can be used to solve the CME model. Finally we will summarize the existing statistical methods that can be used to infer unknown parameters or structures in the CME model using single-cell-level gene expression data.}
\section{Introduction}

In his influential book ``\textit{What is Life}," physicist Erwin Schr\"{o}dinger suggested that macroscopic physical laws often rise from chaos on amicro scale, which was dubbed as "order-from-disorder." (Schr\"{o}edinger, 1944).Schr\"dinger's idea had a strong impact on the development of modern molecular genetics (Dronamraju, 1999) and sheds light on our inquiry about the inner mechanism of biological systems. While living organisms are capable of extracting information from the genome and carrying out complex tasks with great precision, the functions of individual cells are subjected to various fluctuations inherent to the microscopic world. Nonetheless, due to the limitations of experimental techniques, early studies on cellular systems were often carried out based on the ensemble-level data measured over a huge number of cells (with a few notable exceptions, such as the experiment on the induction of the bacterium \textit{E.coli.} conducted by Novick and Weiner,1957). In such ensemble-level experiments, only the average properties of cells in a given population are measured, and the information regarding individual cells is lost. Consequently, many existing mathematical models used for describing biological systems are essentially deterministic. Stochastic elements, if included, often only serve the role of representing external noise. 

In recent years, the development of microscopic measurement technology has enabled scientists to study biological processes at the single-cell or single-molecular level. Notable approaches include single molecule fluorescence in situ hybridization (smFISH), an imaging-based method in which multiple oligonucleotides are used as probes to visualize the quantities and locations of individual molecules within a single cell (Raj, \textit{et al.,} 2008); flow cytometry and mass cytometry, in which fluorescent dyes or heavy-metal-conjugated antibodies are used as markers so that the abundances of multiple types of proteins (18 in flow cytometry and more than 40 in mass cytometry) within the same cell can be measured simultaneously (Chattopadhyay, \textit{et al.,} 2006, Bandura, \textit{et al.,} 2009, Bendall, \textit{et al.,} 2012); and single-cell RNA sequencing (scRNA-seq), in which isolated cells are analyzed using next-generation sequencing technology. This approach, although subjected to larger errors, allows whole genome measurements of mRNA species at the single-cell level  (Tang, \textit{et al.,}  2010, Gr\"{u}n and Oudenaarden, 2015, Bacher and Kendziorski, 2016). Despite of the difference in the measurable quantities, all these methods provide direct experimental evidences on the existence of intrinsic noise: The expressions of genes differ significantly across individual cells, even the cells from the same population.

Generally speaking, the root of ``intrinsic noise" lies in the numerous chemical reactions responsible for the productions and degradations of mRNA and protein molecules. All chemical reactions are the result of collisions between discrete molecules in constant random motions (McQuarrie,1967).The inherent stochasticity in molecular kinetics propagates through the chains of reactions and eventually manifests as observable fluctuations in gene expression. On the ensemble-level, such intrinsic fluctuation is negligible because of the law of large numbers. However, on the single-cell level, the impact of intrinsic fluctuation is much more noticeable due to the so-called ``low-copy-number" effect. First, within a single cell, there are usually only one or two copies of each gene. Second, the copy number of each mRNA specie tends to be small, partly due to the relatively short time span of mRNA molecules (Bernstein, \textit{et al.,} 2008). Third, the copy numbers of many protein species, although they are often relatively large, may still exhibit a high degree of variability due to the fluctuations in mRNA numbers (Ozbudak,\textit{et al.,} 2002; Cai, Friedman, and Xie, 2006). In addition, the copy numbers of certain important protein species serving as transcription factors may also be low (Xie, \textit{et al.,} 2008). Thus, any fluctuation in the number of molecules within a single cell is relatively significant and may lead to important biological consequences. 

In fact, intrinsic noise often functions as an integral part of the control system within a cell. It is known that gene regulatory systems employ complex mechanisms to fulfill essential biological functions. While some of these mechanisms serve as noise suppressors (Becskei and Serrano, 2000), many other mechanisms can only be realized by actively utilizing intrinsic noise. Notable examples include the genetic toggle switch (Shea and Acker, 1985; Ptashne,1986; Gardner, Cantor, and Collins, 2000;), the genetic oscillator (Ishiura \textit{et al.,} 1998; Elowitz and Leibler, 2000) and the frequency-modulation regulation (Cai,  Dalal, and Elowitz, 2008). Through such mechanisms, intrinsic noise enables cells to stochastically switch between different phenotypes and allows homogeneous cell populations to differentiate into inhomogeneous populations (K\ae rn,\textit{et al.,} 2005; S\"{u}el,\textit{et al.,} 2007; Maamar, Raj, and Dubnau, 2007; Lidstrom and Konopka, 2010; Bal\'{a}zsi, van Oudenaarden and Collins, 2011). In particular, it is suggested that intrinsic noise may play a key role in the stem cell differentiation (Hanna, \textit{et al.,} 2009, Yamanaka, 2009; Garg and Sharp 2016) and cancer development (Ao \textit{et al.,} 2008; Wang \textit{et al.,} 2014). Moreover, intrinsic noise may help the cell to adapt to environmental change and provide an evolutionary advantage (Thattai and van Oudenaarden, 2004; Acar, Mettetal, and van Oudenaarden, 2008). 

Therefore, the observed intrinsic noise in gene expression, or more precisely, the empirical distribution of the copy numbers of different mRNA or protein species at the single-cell level, contains valuable information on the regulatory mechanism of the cell. Compared with a study utilizing only the ensemble-level data, the analysis of intrinsic noise would provide not only more detailed information to improve the precision of inferences, but would also unique insight that would not be available from ensemble-level data (Munsky, Fox and Neuert, 2015). It has been shown that a very simple analysis of intrinsic noise could enrich our understanding of regulatory system. For instance, strong correlation coefficients between the expressions of different genes can be used as indicators of the existence of a direct or indirect regulatory relationship (Stewart-Ornstein, Weissman, and El-Samad, 2012). If the single-cell-level gene expression could be monitored over time, the autocorrelation function could be used to determine the direction of regulatory relationships (Dunlop et al, 2008). Under the ergodicity assumption, the snapshots of single-cell-level gene expression could be used to reconstruct the temporal dynamic of a cellular system, such as the decision of cell fates (Trapnell, \textit{et al.,} 2014) and the oscillation of the gene regulation system (Leng, \textit{et al.,} 2015). Furthermore, as the characteristics of intrinsic noise differ between different phenotypes of cells, it has been suggested that the profiles of intrinsic noise should be used alongside with profiles of mean expressions to distinguish phenotypes (Wills, \textit{et al.,} 2013). Nonetheless, in order to fully utilize the information stored in intrinsic noise and to draw quantitative conclusions with clear physical interpretations, we need to develop suitable stochastic models and coupling inference approaches that can account for the key traits of single-cell systems. 

There are many existing models that can be used to study gene regulatory systems, notable examples include logical models (Kauffman, \textit{et al.,}, 2003), linear or nonlinear deterministic models (Klipp  \textit{et al.,} 2008) and Bayesian networks (Friedman,\textit{et al.,} 2000). In this paper, we will focus solely on models and inference methods that are based on or closely related to the chemical master equation (CME). Under this framework, a gene regulatory system consists of a number of different molecular species, as well as a number of chemical reactions in which the molecular species serve as reactants and products. The state of a CME system at a given time is determined by a discrete random vector representing the copy numbers of each molecular species. And the occurrences, or firings, of the chemical reactions modify the copy numbers of the involved species and consequently drive the dynamical evolution of the system state. Under appropriate assumptions, such a system can be modeled as Markov point process. Then the distribution of the system state over time can be described by a set of differential equations collectively known as the CME, a discrete form of the famous Kolmogorov forward equation. CME models boast quite a few advantages for studying a system at the single-cell level. Firstly, they capture the discrete, stochastic and dynamical nature of a cellular system and can offer a casual and physical explanation of the observed data. Secondly, such models are quite flexible and can be used to represent systems with different levels of details. For instance, a simple model can be expanded by incorporating intermediate species and chemical reactions not presented in the original model. It can also serves as a starting point for a more sophisticated model if a certain key assumption, such as the Markov property, has to be dropped. Thirdly, many existing models, including the stochastic differential equation models, deterministic differential equation models and even the logical models, can be viewed as approximations of the CME. For this reason, the CME-based approach allows us to unify different models under a single framework.  

In this paper, we will firstly introduce the concept of the CME and explain how the CME can be used to model a gene regulatory system. We will then discuss several popular alternative approximations of the CME. Afterward we will discuss how to study the dynamics of a given CME system through the means of simulation and through the means of analytically or numerically solving the CME. Finally we will discuss the related inference approaches available in the current literature to conclude our article. 

\section{Modeling with Master Equation}

\subsection{An Overview of the Chemical Master Equation Model}

The first application of the CME in analyzing chemical and biological systems is attributed to Delbr\"{u}ck (1940), who constructed the CME to model the distribution of the number of molecules in autocatalytic reaction $A+B\rightarrow 2B$ (molecular B serves as the catalyst for converting a molecule A to a molecule B). Later, similar methods were used to study bimolecular reactions, such as $A+B\rightarrow C$ (molecule A combines with molecule B to form molecule C,  R\'enyi 1954), $2A\rightarrow B$ (the copies of molecules A forms a single molecule B, Ishida, 1960). In these relatively simple cases, the distribution of the copy numbers of molecular species can be solved analytically, often with the aid of probability generating function (Singer 1953). Thorough reviews of these early works can be found in Bharucha-Reid (1960) and McQuarrie (1967).

Generally speaking, the construction of the CME starts with a system with $I$ different molecular species $S_1, S_2, \cdots, S_I$. Assuming that these molecular species exist in a confined space with a fixed volume and all the molecules are well mixed, then the state of this system only depends on the copy numbers of each molecular species. Let us use $X_i$ to denote the copy number of molecular species $S_i$, then the discrete and non-negative vector $\bold{X}=(X_1, \cdots, X_I)^T$ represents the system state. The evolution of this system, in terms of $\bold{X}$, is driven by various chemical reactions between these species. Generally speaking, a particular chemical reaction can be represented by the following formula:

\begin{equation}b_1S_1+b_2 S_2+\cdots+b_I S_I  \xrightarrow[]{\tau} c_1S_1+c_2 S_2+\cdots+c_I S_I\label{Chemical_Reaction}\end{equation}

\noindent where $b_i$ represents the number of molecule $S_i$ required as reactant, and $c_i$ represents number of molecule $S_i$ in the product. When this reaction fires, $(b_1, b_2, \cdots,b_I)^T$ copies of species $S_1,\cdots,S_I$ will be transformed into $(c_1, c_2,\cdots, c_I)^T$ copies of species $S_1,\cdots,S_I$. Starting from a initial state $\bold{X}$, one firing of this reaction would result in a net change $\boldsymbol{\xi}=\bold{c}-\bold{b}$, and update system state from $\bold{X}$ to $\bold{X}+\boldsymbol{\xi}$. For instance, in the autocatalytic reaction $A+B\rightarrow 2B$, molecule $B$ will serve as catalyst to convert a molecule $A$ to a molecule $B$, and a firing of this reaction would result in a net change $(-1,1)^T$ to the copy numbers of species $A$ and $B$. 

The frequency of the firing depends on the rate constant ${\tau}$, current system state $\bold{X}$ and the volume of the system. As we have discussed in the introduction, all reactions are driven by the random collisions between molecules. Consequently, the higher the number of molecules, the smaller the volume, and the more frequently the reaction fires. If we want to quantify such relationships precisely, we would need the following assumptions:  1) The system evolves within a fixed volume $\Omega$, 2) molecules in the systems are always well-mixed, and 3) the collisions between molecules are sufficiently elastic so that only a small percentage of collisions trigger the firing of the reaction. It can then be shown that, the time between successive firings of a reaction follows an exponential distribution whose rate is proportional to the number of different ways of selecting $(b_1, b_2, \cdots, b_I)^T$ molecules out of $(X_1, \cdots, X_I)^T$ and is inversely proportional to $\Omega^{\sum b_i-1}$. This is equivalent to the counting problem of determining the number ways of a particular color combinations in a given box when balls with different colors are randomly assigned to $\Omega$ different boxes. Thus, the rate of exponential distribution would roughly be proportional to $\prod_{i=1}^I X_i^{b_i}  / \Omega^{\sum b_i-1}$ (Gillespie, 1992). If we use $\tau$ to denote the proportional rate constant and set $\Omega=1$ for simplicity, the rate of the firing of chemical reaction (\ref{Chemical_Reaction}) is

\begin{equation}a(\bold{X})=\tau \prod_{i=1}^I X_i^{b_i}\label{Propensity_Func}\end{equation}

\noindent which is also known as the propensity function \footnote{Another commonly used formulation of propensity function is $a(\bold{X})=\tau \prod_{i=1}^I X_i ! / (X_i-b_i)!$. This formulation further emphasizes the discrete nature of molecular system. For instance, the number of choosing two molecules from $n$ molecules of the same type should be $n(n-1)$ instead of $n^2$.}. When multiple reactions are present, the waiting time for each reaction to fire is independently distributed with its own propensity function. The reaction that fires first may change the system state and consequently modify the propensity function of other reactions, however. 

The memoryless property of the waiting time between successive reactions implies that the evolution of such a system can be modeled as a point Markov process. From now on, we will use notation $\bold{X}(t)$ to specify the system state at time $t$ (we may omit the time indicator for simplicity whenever possible). Let us assume that there are $K$ different reactions in the system. Denote the propensity function of the $k$th reaction as $a_k(\bold{X})$, which should only depend on time $t$ through $\bold{X}$, and the net change induced by one firing of the $k$th reaction as $\boldsymbol{\xi}_k$; then the probability of finding the system in a particular state $\bold{X}$ at time $t$ obeys the CME:

\begin{equation} \frac{d \P_t(\bold{X})}{dt}= -[\sum_{k=1}^K a_k(\bold{X})] \P_t(\bold{X}) +  \sum_{k=1}^K a_k(\bold{X}-\boldsymbol{\xi}_k) \P_t(\bold{X}-\boldsymbol{\xi}_k),\label{CME}
\end{equation}

\noindent a differential equation of $\P_t(\bold{X})$ (Van Kampen, 1992). Here the time derivative of the distribution of the system is decomposed into the loss of probability mass as the system moves out state $\bold{X}$ and the gain of probability mass as the system moves into state $\bold{X}$. These two components are further represented by the summation of contributions from individual reactions. 

As long as the initial distribution $\P_0(\bold{X})$ is specified, the distribution $\P_t(\bold{X})$ at any later time $t$ will be determined by CME. And if we set the time derivative to 0 and solve equaltion (\ref{CME}), we will obtain the equilibrium (steady state) distribution $\P^s(\bold{X})$. This solution is unique and stable in the sense that, as $t\rightarrow\infty$, $\P_t(\bold{X})$ will always converge to  $\P^s(\bold{X})$ as long as $\P_0(\bold{X})$ is a proper probability vector (Schnakenberg, 1976).

A CME system whose propensity functions are all in the form of equation (\ref{Propensity_Func}) is also known as a stoichiometric network. Mathematically it is also feasible to construct a CME with more complicated propensity function. One popular choice in the chemical and biological literature is to use the rational function;  we will discuss some examples later.

Equivalently, the stochastic process $\bold{X}(t)$  can also be represented as the solution to the following stochastic equation (Ball, \textit{et al.,} 2006; Anderson 2007):

\begin{equation} \bold{X}(t)= \bold{X}(0)+\sum_{k=1}^K Y_k [\int_0^t a_k(\bold{X} (s)) ds]\boldsymbol{\xi}_k, \label{PoissonRep}
\end{equation}

\noindent in which $Y_k(t)$ represent independent and unit-rate Poisson Processes, and the term $Y_k [\int_0^t a_k(\bold{X} (s)) ds]$ represents the total number of firings of the $k$th reaction during period $[0,t]$. For instance, in a pure birth process with only one species, which is created one-copy-at-a-time in a reaction with a constant propensity function $\tau$, the solution to (\ref{PoissonRep}) at time $t$ will be Poisson distribution with rate $\tau t$. One key advantage of representation is that it can be extended relatively easily to non-Markov system, such as system with time delay or with time dependent propensity functions (Anderson 2007); 

It is often impossible to find analytical solution to a CME when the dimension of system state is higher than one. Alternatively, we may focus on the evolution of the moment of $\bold{X}$. In particular, for any function $H(\bold{X})$, its expectation $ \E_t[H(\bold{X})]$ with respect to $\P_t(\bold{X})$ obeys the following equation (Van Kampen, 1992): 
\begin{equation} \frac{d \E_t[H(\bold{X})]}{dt}= \E_t[(H(\bold{X}+\boldsymbol{\xi}_k )-H(\bold{X}))a_k(\bold{X})],\label{moment}\end{equation}
\noindent which can sometimes be used to solve the moment of $H(\bold{X})$ at any time $t$ or at equilibrium. 

While a CME model can be used to describe any chemical and biological system in principle, its application is often limited by its complexity. The formulation of the CME, as we have discussed above, requires that the states of every species be accounted for. This can lead to serious analytical difficulty when the number of species is large and may also be unnecessary in practice. First, scientists may only be interested in the evolution of a few key species even if the system under investigation involves a large number of other species. Second, complete information on the states of every species in the system is usually unobtainable, and the information we can collect from an experiment may not be sufficient for us to identify and infer the parameters in a complete model  (Azeloglu and Iyengar, 2015). Therefore, it is thus of great importance to investigate approaches that can be used to construct a "reduced" version of the CME that only focuses on the key species of interest. 

In their seminal work on the kinetic of enzymatic reaction, Michaelis and Menten (1913) removed the intermediate species in the deterministic differential equation model based on the phenomenon of ``time scale separation". In many chemical and biological systems, the intermediate species may evolve at a much faster rate compared with the other main species under investigation. Consequently, it is then reasonable to propose that such species will always reaches a ``quasi" stable state almost instantly relatively to the main species. Then the approximated abundance of the intermediate species can be determined based the current abundance of main species. This approach, commonly known as qauasi-steady-state approximation (QSSA) or pseudo-steady-state assumption, can be justified using perturbation theory and has been widely used to simplify deterministic biochemical models involving multiple species (Ackers, Johnson, and Shea, 1982; Segel and Slemrod, 1989; Gunawardena, 2014). 

Rao and Arkin (2003) first applied QSSA to reduce the complexity of a CME based model. In their work, the system state vector is partitioned as $\bold{X}=(\bold{Y},\bold{Z})^T$, representing the main species and intermediate species respectively. Then we have:

\begin{align} \frac{d \P_t(\bold{X})}{dt}=  \frac{d \P_t(\bold{Z}|\bold{Y})}{dt} \P_t(\bold{Y})+\frac{d \P_t(\bold{Y})}{dt} \P_t(\bold{Z}|\bold{Y}) \label{QSSA_partition}
\end{align}

The QSSA can then be expressed as follows: conditioning on the main species $\bold{Y}$,  intermediate species $\bold{Z}$ still evolves as a Markov process and reaches the equilibrium distribution infinitely fast. Thus, the time-dependent conditional distribution of $\bold{Z}$ given $\bold{Y}$ can be replaced by time-independent ``conditional equilibrium distribution" in which $\bold{Y}$ is treated as constant. That is, $\P_t(\bold{Z}|\bold{Y})\approx \P^s(\bold{Z}|\bold{Y})$ and $d \P_t(\bold{Z}|\bold{Y})/dt\approx 0$. Then the above equation is simplified as: 


\begin{align*} \frac{d \P_t(\bold{X})}{dt} \approx \frac{d \P_t(\bold{Y})}{dt} \P^s(\bold{Z}|\bold{Y}).
\end{align*}

A new reduced CME can then be used to model the distribution of main species $\P_t(\bold{Y})$. This new CME will only take into consideration of the reactions that result in a net change of $\bold{Y}$. In the following discussion, we will use notation $k^*$ as index for such reaction and use $\boldsymbol{\xi}^*_{k}$ to represent the net change of $\bold{Y}$ due to the firing of the $k$th reaction. Still, the propensity function of such reaction under consideration $a_{k^*} (\bold{Y}, \bold{Z})$, may still depend on intermediate species $\bold{Z}$. This issue can be resolved by replacing it with expected propensity function with respect to $\P^s(\bold{Z}|\bold{Y})$ or simply substituting $\bold{Z}$ with $\E^s(\bold{Z}|\bold{Y})$ (which is essentially a deterministic approximation). Then if we denote  $\bar{a}_k( \bold{Y})=\E^s [a_k (\bold{Y}, \bold{Z}) | \bold{Y}]$, and $\hat{a}_k( \bold{Y})=a_k (\bold{Y}, \E^s [\bold{Z} | \bold{Y}] ) \approx \bar{a}_k( \bold{Y})$, the reduced CME that approximates the evolution of $\bold{Y}$ can be written as:

\begin{align} \frac{d \P_t(\bold{Y})}{dt} & \approx -\sum_{k^*}  \bar{a}_{k^*} ( \bold{Y})\P_t(\bold{Y}) +  \sum_{k^*} \bar{a}_{k^*} (\bold{Y}-\boldsymbol{\xi}^*_{k^*}) \P_t(\bold{Y}-\boldsymbol{\xi}^*_{k^*}) \\ &
\approx -\sum_{k^*}  \hat{a}_{k^*} ( \bold{Y})\P_t(\bold{Y}) +  \sum_{k^*} \hat{a}_{k^*} (\bold{Y}-\boldsymbol{\xi}^*_{k^*}) \P_t(\bold{Y}-\boldsymbol{\xi}^*_{k^*}), \label{QSSA_discrete}
\end{align}

\noindent Under QSSA, given the copy number of main species $\bold{Y}$, it is usually straightforward to determine $\E^s [\bold{Z} | \bold{Y}]$ using a deterministic approximation.  More rigorous theoretical treatment of QSSA can be found in the work of Ball, \textit{et al.,} (2006) and Kang and Kurtz (2013).\newline

\subsection{Chemical Master Equation Model and Gene Regulation System}
In order to model a gene regulation system with the CME, the interactions between genes, mRNAs and proteins should be modeled as a number of reactions in the form of equation (\ref{Chemical_Reaction}). Let us start with the most simplified scenario: a system with a single gene and no regulatory mechanism. The key interactions in this system, as summarized in the central dogma of cell biology, can be represented as four basic reactions: 1) the transcription process, in which an mRNA molecule is produced from the gene; 2) the translation process, in which a protein molecule is produced from one of the existing mRNA molecules; 3) and 4) degradation of the mRNA and protein, in which an mRNA or protein molecule degrades into smaller molecules that may be used for other biological processes. If we use  $``R"$ and $``P"$ to denote mRNA and protein species, respectively, then the mechanism of the system can be represented as follows:

\begin{exmp}[Simple Transcription Model]
\begin{equation*}
\begin{array}{lrcl}
\text{Transcription: } & \text{Gene}  &  \xrightarrow[]{\tau_{R}}  & R +  \text{Gene} \\
\text{Translation: } & R &  \xrightarrow[]{\tau_{P}}   &R+P\\
\text{mRNA Degradation: } & R   & \xrightarrow[]{\lambda_R} &\emptyset \\
\text{Protein Degradation: } & P&\xrightarrow[]{\lambda_P} & \emptyset \\
\end{array} 
\end{equation*}
\label{Model_Dogma}
\noindent where $\emptyset$ is used to denote the products of degradation processes (Thattai and van Oudenaarden, 2001).
\end{exmp}

Assuming that the number of gene within a cell is a fixed constant (thus avoiding the complexity due to cell division), we will only focus on the copy numbers of on mRNA and protein species. Denote the copy numbers as $(X_R, X_P)^T$, then the propensity functions and the induced net changes of the four reactions in Model \ref{Model_Dogma} are:

\[
\begin{array}{lcc}
& \text{Propensity Function} & \text{Net Change} \\
\text{Transcription: } & \tau_R  &  (1,0)^T \\
\text{Translation: } & \tau_R X_R  &  (0,1)^T \\
\text{mRNA Degradation: } & \lambda_R X_R  &  (-1,0)^T \\
\text{Protein Degradation: }& \lambda_P X_P &  (0,-1)^T \\
\end{array} \]
\newline

Then based on equation (\ref{CME}), the CME of this simplified system can be written as following: 

\begin{align*} \frac{d \P_t(X_R, X_P)}{dt}= & -(\tau_R+X_R\lambda_R+X_R \tau_P+X_P\lambda_P) \P_t(X_R, X_P) \\ 
& + \tau_R \P_t(X_R-1, X_P)+  (X_R+1)  \lambda_R \P_t(X_R+1, X_P) \\ & +  X_R  \tau_P \P_t(X_R, X_P-1)+ (X_P+1) \lambda_P \P_t(X_R, X_P+1).
\end{align*}

In principle, the solution of the above equation (given suitable initial conditions) will allow us to analyze the distribution of the number of mRNA and protein at any later time. Unfortunately, no analytical solution exists for either $\P_t(X_R, X_P)$ at time $t$ or the equilibrium distribution $\P^s(X_R, X_P)$ (Shahrezaei and Swain, 2008). Still, as the production and degradation of mRNA do not depend on the number of protein molecule, the evolution of mRNA species alone forms a simple constant-rate birth-death process and can be described using a one-dimensional CME. It is easy to see that the equilibrium distribution of mRNA molecules is a Poisson distribution with rate $\tau_R/\lambda_R$. Furthermore, based on equation (\ref{moment}), differential equations of certain moments, including means, variances and covariances of the copy numbers of mRNA and protein, can be derived and solved (Thattai and van Oudenaarden, 2001; Munsky, B. Trinh, and M. Khammash, 2008). For instance, it can be shown that, at the equilibrium state, the expectation of the number of protein molecule equals $(\tau_R\tau_P)/(\lambda_R \lambda_P)$, and the corresponding variance equals the expectation multiplied by $1+\tau_P/(\lambda_P+\lambda_R)$. That is, the fluctuation in the number of mRNAs propagates through the translation process and amplifies the fluctuation in the number of proteins. One way to utilize this simple result is to calculate the Fano factor, defined as the ratio between variance and mean. If the observed Fano factor of mRNA molecules is significantly larger than 1, the underlying system must be more sophisticated than Model  \ref{Model_Dogma} (Raser and O'Shea, 2004). 

Another implication of Model \ref{Model_Dogma} is that if the transcription rate $\tau_R$ is low while the transcription rate $\tau_P$ is high, the protein molecules will be produced in a "burst" following the transcription of every new mRNA molecule. This phenomenon has been observed and studied extensively in single-cell experiments (Cai, Friedman, and Xie, 2006; Yu, \textit{et al.}, 2006).
\newline

\begin{figure}[htbp]
\begin{center}
\includegraphics[width=0.7\textwidth]{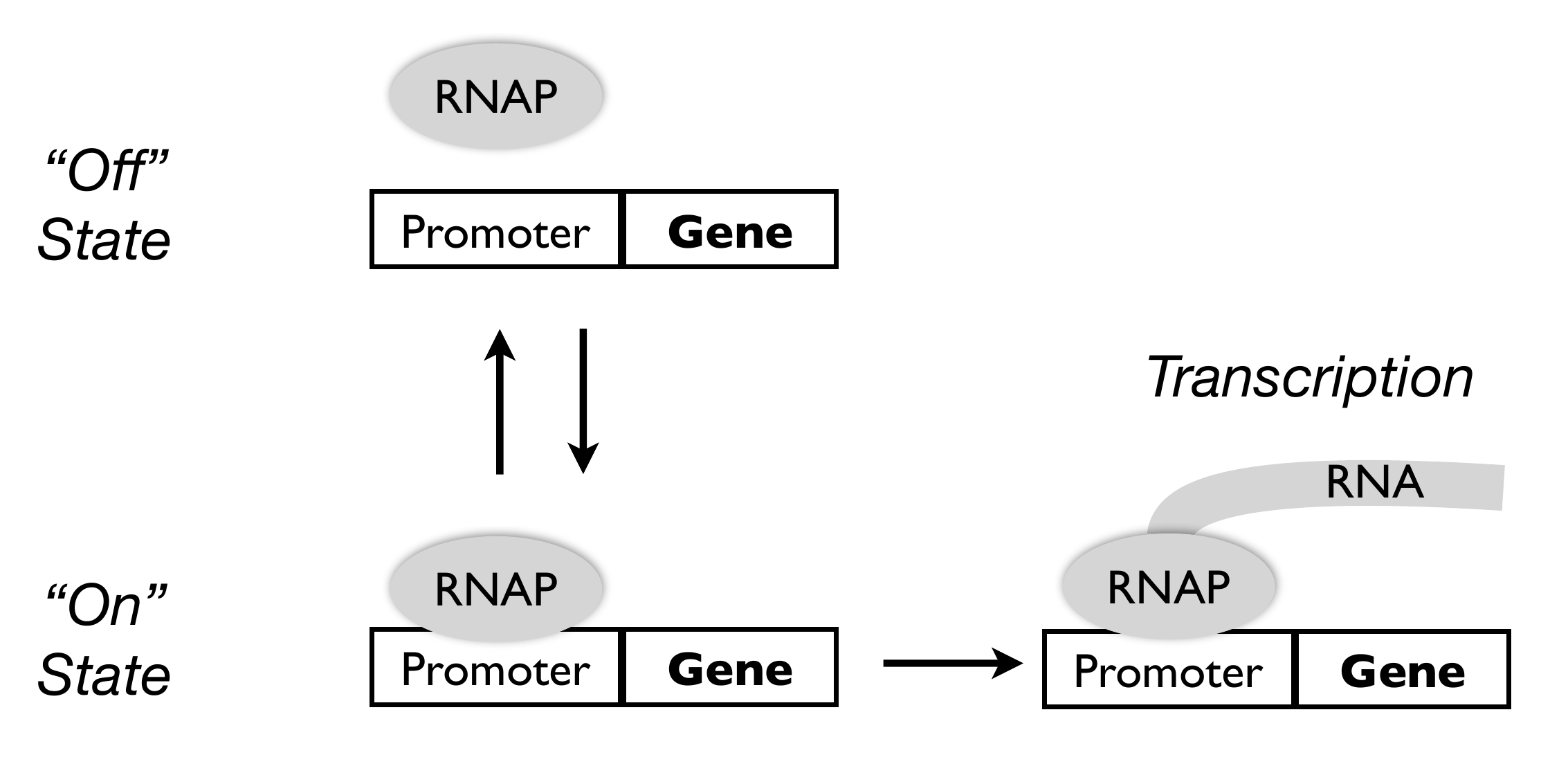}
\caption{Schematic Representation of the Two-state Transcription Model}
\label{2S_Model_figure}
\end{center}
\end{figure}

In order to take into account the more sophisticated mechanisms in gene regulatory systems, we need to give a more detailed account of the transcription process. Generally speaking, to initiate the transcription process, an enzyme called RNA polymerase (RNAP) must attach to a specific site (promoter) near the gene. The structure of chromatin that enclosed the corresponding gene must also go through a process known as chromatin remodeling to allow RNAP to access the promoter site and start the transcription process. One way to account for this fact is to treat the gene itself as a dynamical system. In particular, we may assume that a given gene can stochastically switch between an inactive state and an active state (we will use the notations ``on" and ``off" to distinguish these two states) and that transcription can only occur when the gene is active. If we further assume that all the other factors (such as the number of RNAP molecules) related to the switching behavior can be treated as constant, the transcription process in Model \ref{Model_Dogma} can be expanded into the following two-state transcriptional model (Kepler and Elston, 2001, also see Figure \ref{2S_Model_figure}):

\begin{exmp}[Two-State Transcriptional Model]
\begin{equation*}
\begin{array}{rcl}
 \text{Gene}_{\text{off}} & \xrightleftharpoons[k_{\textit{off}}]{\,k_{\textit{on}}\,} & \text{Gene}_{\text{on}} \\
 \text{Gene}_{\text{on}}  & \xrightarrow[]{\tau_{R}}  & R+ \text{Gene}_{\text{on}} \\
\end{array} 
\end{equation*}
\noindent in which the first line represented two coupled reactions in opposite directions. 
\label{2S_Model}
\end{exmp}

The state space of this new model should be expanded to account for the state of the gene. For instance, the inactive and active states of the gene may be coded as ``0" and ``1", respectively. Under this model, the steady state distribution of the mRNA molecule is no longer a Poisson distribution but can still be obtained analytically (Raj, \textit{et al.,} 2006).

%
%
%

When the rate of switching between ``active" and ``inactive" states is high, the qualitative behavior of this two-states model will be quite similar to the behavior of Model \ref{Model_Dogma}. However, slow switch rates will generally result in a transcriptional ``burst" of mRNA molecules, and increase the cell-to-cell variability of mRNA numbers. Consequently, the tail of the mRNA copy number distribution will be longer than the tail of a Poisson distribution, and bi-modality may appear if the probability of the gene being inactive is significant (Kepler and Elston, 2001; Shahrezaei and Swain, 2008). For the experiment evidences, readers can refer to the work of Golding, \textit{et al.,} (2005) and Raj, \textit{et al.,} (2006).


Due to the complexity of the transcriptional process, this aforementioned two-state model has been expanded into a multiple-state model to incorporate a sophisticated control mechanism. For instance, to describe certain features specific to eukaryotic transcription, a five-states stochastic model is used to study eukaryotic gene expression (Blake, \textit{et al.}, 2003). The core system that controls the life cycle of virus \textit{Lambda phage} contains 5 different promoter regions and may take 32 different states (Lei, \textit{et al.,}2015). As demonstrated in the work of Neuert, \textit{et al.,} (2013), in order to fit the observed mRNA distribution of STL1 gene in \textit{Saccharomyces cerevisiae}, a four-state regulation model is used. It is safe to predict that more sophisticated models will be needed to explain the data emerging from the experiments that employ more advanced technology.  Still, a CME-based approach, in principle, can be easily modified to reflect the newly observed phenomena. 
\newline

The discussion above focuses on a single gene, but it is quite straightforward to expand both Model \ref{Model_Dogma} and Model \ref{2S_Model} to accommodate multiple genes that act independently, with their own mRNA and protein species. In the subsequent discussion, numerical subscripts $1, 2, \cdots$ will be used to label different species of genes, mRNA and proteins. The difficult part, however, is to incorporate the regulatory interactions between genes. We will focus on how to expand model \ref{2S_Model} to account one of the most common regulatory mechanisms: the transcriptional regulation. 

Within a cell, proteins of the same or different species often form complexes known as transcriptional factors (TF) that serve as regulators in the transcriptional process. For instance, a transcriptional factor may block the binding of RNAP molecules to the promotor region and thus prevent the initiation of transcription. Another transcription factor might serve to improve the affiliation between RNAP molecules and promoter region, increasing transcriptional activity. Through transcriptional factors, complex regulatory networks can be formed between genes. In practice, a complex gene regulation network is often formed by many small sub-units with distinctive characteristics, which are called motifs or modules. Here we will use one of the most commonly studied regulatory motifs, which was firstly used to model the control system of bacterial known as \textit{Enterobacteria phage }$\lambda$ (Shea and Acker, 1985; Arkin, Ross and McAdams, 1998) as an example to illustrate how a complex transcription regulation mechanism can be modeled with the CME. 

\begin{figure}[htbp]
\begin{center}
\includegraphics[width=0.7\textwidth]{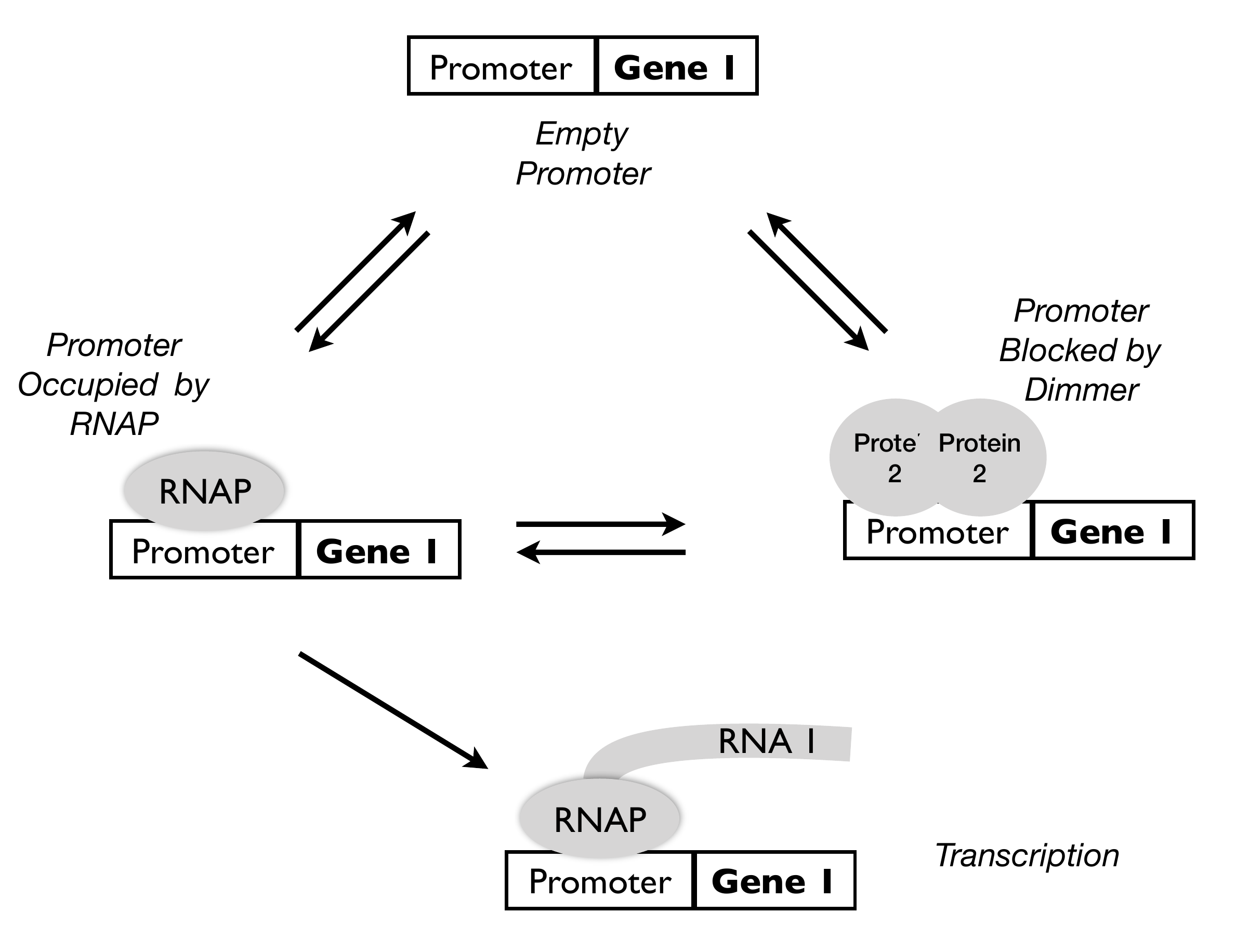}
\caption{Schematic Representation of the Regulated Transcription Model}
\label{Repressor_Model_figure}
\end{center}
\end{figure}

This motif contains two genes and their corresponding mRNA and protein species.  The transcription activity of each genes is repressed by the transcription factor formed by the protein molecules of another gene. Let us focus on the transcription process of the first gene. This process can be summarized in the following model (also see Figure \ref{Repressor_Model_figure} for schematic representation):

\begin{exmp}[Repressor Model]
\begin{equation*}
\begin{array}{rcl}
 2P_2 & \xrightleftharpoons[k_{-1}]{\,k_{1}\,} & P_2\text{-}P_2 \\
P_2\text{-}P_2+  \text{Gene}_{1,\text{off}} & \xrightleftharpoons[k_{-2}]{\,k_{2}\,} & \text{Gene}_{\text{1}}\text{-}P_2\text{-}P_2\\
 \text{Gene}_{1,\text{off}} & \xrightleftharpoons[k_{\textit{off}}]{\,k_{\textit{on}}\,} & \text{Gene}_{1,\text{on}} \\
 \text{Gene}_{1,\text{on}}  &  \xrightarrow[]{\tau_{R_1}}  & R_1 \\
\end{array}
\end{equation*}
\label{Repressor_Model}
\end{exmp}

Similar to the previous models, the transcription process of the first gene can be initiated if the first gene is in active state. However, two copies of the protein molecules of the second gene $(P_2)$ can form a dimer ($P_2\text{-}P_2$). This dimer can interact with the first in its inactive state and prevent it from switching to its active state. Therefore, an increase in the copy number of $P_2$ will lead to an increase in the number of dimer $P_2\text{-}P_2$. Then the first gene in inactive state will be more likely to bind with a dimer rather than switching into active state and initiating transcription. Therefore, an increase in the abundance of molecules $P_2$ will eventually results in an overall decrease in the transcriptional activities of the first gene. That is, the second gene represses the expression of the first gene through its protein products. More specifically, as the first gene can now stochastically switch between three mutually exclusive states (inactive state $\text{Gene}_\text{1,off}$, binding state $ \text{Gene}_{\text{1}}\text{-}P_2\text{-}P_2$, and active state  $\text{Gene}_\text{1,on}$), and the abundance of $P_2$ increases the relative weight of the binding state, the repression mechanism discussed here is nonlinear rather than linear. (Shea and Acker, 1985).

If the first gene also represses the expression of the second gene in a similar fashion, then the equilibrium distribution of the system described above may contain two distinctive modes with the choice of appropriate parameters (Cherry and Adler, 2000; Kepler and Elston, 2001; Tian and Burrage 2006; Smadbeck and Kaznessis, 2013).  In each mode, the expression of one of the genes will be high and the expression of the other gene will be suppressed. Switching between modes might occur stochastically when the intrinsic fluctuation is sufficiently strong to drive the system across the barrier between two modes. When the barrier between two modes is high, the waiting time between switches might be so long that the system will effectively settle in one of the modes unless an external stimulus is applied. This system, commonly known as a genetic switch in the biology literature, represents one of the most commonly motifs found in real biological regulatory systems and has been artificially engineered in the lab (Ptashne, 1992; Gardner, Cantor, and Collins, 2000). 
\newline

All three of the models discussed above contains propensity functions in the form of equation \ref{Propensity_Func} and thus they all belong to the family of stoichiometric networks. While these examples illustrate how the complex nonlinear regulatory mechanisms can be incorporated in a CME-based framework as a set of chemical reactions, such formulation may require the tracking of a large number of intermediate species. For instance, in model \ref{Repressor_Model}, the state vector of the CME should include the following variables: number of second protein molecules, number of dimers formed by the second protein, number of the first mRNA molecules and current state of the first gene. It is unrealistic to imagine that we could obtain all of this information in a single experiment. A common approach to handling such a situation is to apply the QSSA to simplify the transcriptional regulation model through removing the intermediate species.  

Let us take Model \ref{Repressor_Model} as an example. Suppose that we treat the dimer and the three different states of the first gene as intermediate species. Under QSSA, given the number of main species (mRNA and proteins), the distributions of these intermediate species always attain the corresponding conditional equilibrium distributions. We can then further approximate the number of dimmer with its conditional expectation and the indicator random variables that denoting the states of first gene with their conditional probabilities. The quasi-equilibrium condition will demand that the propensity functions in each of the three reversible reactions of \ref{Repressor_Model} to be equal on average. Then we have the following equalities:

\begin{equation}
\begin{array}{rcl}
 k_1 X_{P_2}^2 & = & k_{-1} \E^s(X_{P_2\text{-}P_2}) \\
k_2 \E^s(X_{P_2\text{-}P_2}) \P^s_{\text{Gene}_{1,\text{off}}} & =& k_{-2} \P^s_{\text{Gene}_{\text{1}}\text{-}P_2\text{-}P_2}\\
k_{\textit{on}}\P^s_{\text{Gene}_{1,\text{off}}} & =& k_{\textit{off}} \P^s_{\text{Gene}_{1,\text{on}}}, \\
\end{array}\label{Repressor_Model_QSSA}
\end{equation}

\noindent which would allow us to derive the probability that the first gene is active as a function of $X_{P_2}$:

\begin{equation} \P^s_{\text{Gene}_{1,\text{on}}} =F_1(X_{P_2})=\frac{1}{1+\frac{k_{\textit{off}}}{k_{\textit{on}}}+\frac{k_{\textit{off}}}{k_{\textit{on}}} \frac{k_{2}}{k_{-2}}\frac{k_{1}}{k_{-1}}  X_{P_2}^2 }=\frac{b_0}{1+c_1 X_{P_2}^2}\label{rational_func}. \end{equation}



\noindent  As the transcription can only be initiated when the gene is active, the average transcriptional rate can be viewed as the product of the transcription rate when the gene is active and the probability that the gene is active. Therefore, instead of modeling the detailed regulatory relationship as in model \ref{Repressor_Model}, we can simplify the model by imposing a hypothetical one-step reaction in which the reaction rate is defined as the average transcription rate:

\begin{exmp}[Repressor Model with Rational Propensity Function]
\begin{equation*}
\begin{array}{rcl}
 \text{Gene}_{1} \xrightarrow[]{\tau_{R_1} F_1(X_{P_2})}  R_1+ \text{Gene}_{1}, &
 F_1(X_{P_2})=\frac{b_0}{1+c_1 X_{P_2}^2}
 \end{array}
\end{equation*}
\label{Nonlinear_Repression}
\end{exmp}

Unlike the multiplicity term in equation (\ref{Propensity_Func}), the propensity function of the above reaction is a rational function. It is possible to use a similar rational function to capture the nonlinearity of the more complex regulatory relationship. This formulation can also be derived based on thermodynamic theory given an appropriate equilibrium assumption (Keller, 1995; Bintu, \textit{et al.,} 2005a, 2005b). Roughly speaking, as we have discussed before,  transcriptional regulation is achieved through interactions between transcriptional factors, RNAP molecules and different binding sites along genes. Different arrangements of these elements can result in different conformations of the system (see the three conformations in Figure \ref{Repressor_Model_figure}) that compete against each other. Transcription can only be initiated when the system enter a ``transcriptional-friendly" conformation that allows a proper binding of RNAP to the promoter site. Thus, the average transcription rate is proportional to the probability that the system is in ``transcriptional-friendly" conformations. According to the thermodynamic theory, this probability can be calculated as the ratio of the weighted number of ``transcriptional-friendly" conformations to the weighted number of all possible conformations. The number of a certain conformation is proportional to the number of corresponding transcriptional factors in the system, which can be determined based on the current abundance of protein molecules and is reflected by the individual terms of polynomial functions in both the numerator and denominator of the ratio. A high abundance of a certain type of molecules tends to increase the prominence of the corresponding regulatory mechanism. The weight of the number of conformations, represented as the coefficients in the rational function, is determined based on the energy associated with a particular type of conformation. Such coefficients would eventually determine the role of the corresponding transcriptional factor. In particular, if a transcriptional factor leads to a conformation that makes RNAP binding more difficult, it will serve as a repressor and the rational propensity function of the transcription rate will be a decreasing function with respect to the abundance of this transcriptional factor. On the other hand, if a transcriptional factor leads to a conformation that reduces the energy required for RNAP binding, it will serve as an activator. We will use the following illustrated example to show that how a given nonlinear rational function can be used to describe a complex regulatory relationship. More general discussion can also be found in Meister \textit{et al}. (2013).

Unlike the multiplicity term, the propensity function of the above reaction is a rational function. It is possible to use a similar rational function to capture the nonlinearity of the more complex regulatory relationship. SuchThis formulation can also be derived based on thermodynamic theory given an appropriate equilibrium assumption. Roughly speaking, as we have discussed before, the transcriptional regulation is achieved through the interactions between transcriptional factors, RNAP molecules and different binding sites along the genes. Different arrangements of these elements can result in different conformations of the system that compete against each other. And transcriptionTranscription can only be initiated when the system enters a ``transcritiohal-friendly" conformation that enables the proper binding of RNAP to the promoter site. Thus, the average transcription rate is proportional to the probability that the system is in ``transcritiohal-friendly" conformations. According to the thermodynamic theory, this probability can be calculated as the ratio of the weighted number of ``transcriptional-friendly" conformations to the weighted number of all possible conformations. The number of a certain conformation is proportional to the number of corresponding transcriptional factors in the system, which can be determined based on the current abundance of protein molecules and is reflected by the individual term in polynomial functions in both numerator and denominator of the ratio. Then a. A high abundance of a certain type of molecules would tendtens to increase the prominence of the corresponding regulatory mechanism. The weight of the number of conformations, represented as the coefficients in the rational function, areis determined based on the energy associated with a particular type of conformation. And suchSuch coefficients would eventually determine the role of the corresponding transcriptional factor. In particular, if a transcriptional factor leads to a conformation that makes RNAP binding more difficult, the rational function wouldwill be a decreasing function with respect to the abundance of this transcriptional factor. On the other hand, if a transcriptional factor leads to a conformation that reduces the energy required for RNAP binding, it would serveswill serve as an activator. We will use the following illustrated example to show that how a given nonlinear rational function can be used to describe a complex regulatory relationship. More general discussion can also be found in Meister \textit{et al}. (2013).

$$\tau_{R_1} F_1(X_{P_1}, X_{P_2}, X_{P_3})=\tau_{R_1} \frac{0.5+X_{P_1}+0.1 X_{P_2} X_{P_3} }{1+X_{P_1}+X_{P_2}^2 +0.4 X_{P_2} X_{P_3}}.$$

\noindent Specifically, in this equation, the ratio of constants in the numerator and denominator represent a base probability of initiating the transcriptional process when all the regulators are absent. The other algebraic terms contained in the equation suggest that there are three different TFs that serve as the regulators of the first gene:  $X_{P_1}$ represents $P_1$, the protein molecule of the first gene;  $X_{P_2}^2$ represents $P_2\text{-}P_2$,  a dimer formed by two copies of protein molecules of the second gene; and  $X_{P_2} X_{P_3}$ represents $P_2\text{-}P_3$, a complex formed by each copy of protein molecules of the second and third genes. When a term (such as $X_{P_2}^2$) only appears in the denominator, it suggests that the corresponding TF, if it binds with the gene, will completely block the transcriptional process. Thus, this TF will tend to slow down the overall transcriptional rate and serve the role of repressor. If a term appears in both the numerator and denominator (such as $X_{P_1}$ and $X_{P_2} X_{P_3}$), the role of the corresponding TF will be determined based on the comparison of the ratio of its coefficients in the numerator and denominator with the base probability of initiating the transcription process. For instance, in this example, $P_1$ will serve as an activator that increases the transcriptional rate, but $P_2\text{-}P_3$ will serve as a repressor. Finally, due to the restriction posed by the thermodynamic theory, all the coefficients should be non-negative, and the coefficient of a given algebraic term in the numerator should be no larger than the corresponding coefficient in the denominator. 

In summary, the application of the nonlinear rational function for describing complex biological systems has two major advantages:  1) It does not require the intermediate species and will solely focus on major species of interest, and  2) it still retains the nonlinear nature of the biological system. Apart from its earlier applications in chemical reaction kinetics such as Michaelis-Menten kinetic (Michaelis and Menten, 1913) and Hill functions (Hill, 1913), it has been used to study gene regulatory systems as well, including negative feedback loops that reduce the intrinsic noise (Paulsson and Ehrenberg, 2000; Becskei and Serrano, 2000; Thattai and van Oudenaarden, 2001), positive feedback loop that may serve as noise amplifiers or as switches (Hasty \textit{et al.,} 2000; Isaacs \textit{et al.,} 2003; Maamar, Raj, and Dubnau, 2007;), and enzymatic reactions (Qian 2008) as well as the dynamica of stem cell switches (Chickarmane, et al., 2006; Chickarmane and Peterson 2008).
\newline

Finally, in the experiment where only the mRNA species or protein species are observable at the single-cell level, we may need to simplify the aforementioned model further so it would only includes the observable species. As the translation process is treated as a simple birth process with a constant rate, we may apply QSSA and state that the copy number of mRNA molecule is proportional to the copy number of the corresponding protein molecule. In this way, we could propose the a simplified model that consists of only mRNA or protein species. For a system with $M$ different genes, we could then use $S_i$ to represent the $i$th species (which could be mRNA or protein species) with copy number $X_i$. Then we could represent the regulation system with the following reactions:

\begin{exmp}[Multple-Gene Model with Rational Propensity Function]

\begin{equation*}
\begin{array}{rcl}
& \cdots &\\
 \text{Gene}_i  &  \xrightarrow[]{\tau_{i} F_i(\bold{X})}  & S_i + \text{Gene}_i \\
S_i&  \xrightarrow[]{\lambda_i}  & \emptyset \\
& \cdots &\\
\end{array}
\end{equation*}
\noindent in which $F_i(\bold{X})$ should be a nonlinear rational function of $\bold{X}=(X_1, X_2, \cdots, X_M)^T$.
\label{Simplified_Rational_Model}
\end {exmp}

Correspondingly, the CME should be:

\begin{align*} \frac{d \P_t(\bold{X})}{dt}= & -[\sum_{i=1}^M (\tau_{i} F_i(\bold{X})+\lambda_i)] \P_t(\bold{X}) \\ 
& + \sum_{i=1}^M \lambda_i (X_i+1) \P_t(\bold{X}+\boldsymbol{\epsilon}_i) \\ & + \sum_{i=1}^M \tau_{i} F_i(\bold{X}-\boldsymbol{\epsilon}_i) \P_t(\bold{X}-\boldsymbol{\epsilon}_i)
\end{align*}

\noindent in which $\boldsymbol{\epsilon}_i$ represents the $K$ dimensional unit vector whose $i$th component equals 1. \newline

In the above discussion, we outline the general idea of applying the CME to model a gene regulatory system. This approach would allow us to capture the discrete, stochastic and nonlinear traits of a biological system at the single-cell or single-molecule level, and it is also flexible enough for expansion. In the following sections, we will outline the general approaches that are available for studying CME-based systems, especially in the context of gene regulation systems. In particular, we will discuss how to approximate, simulate and solve a CME, as well as relevant statistical inference methods for studying a CME-based model. 
\section{Approximation to the Chemical Master Equation}

Under appropriate conditions, it is possible to approximate the CME with a continuous stochastic process, or a deterministic process. In this section we will discuss several approximated schemes of the CME. We will focus on their connections to the unmodified CME, as well as their relative strengths and limitations. 

The various schemes we will discuss here can often be viewed as approximations to the CME under the macroscopic limit. Here,``macroscopic limit" means the limit behavior of a CME system as its size expands to infinity. In our previous discussion, we assumed that a CME-based system evolves within a fixed unit volume (such as the gene regulatory system within a single cell). Then, a system with volume $\Omega$ can be created by pooling $\Omega$ identical copies of the unit-volumed system together. Before merging, these ``identical copies" should be viewed as independent systems that evolved under the same CME. After merging, we will assume that all the particles can travel freely in the expanded system and mix instantly. We will denote the state of the expanded system as  $\bold{X}^{(\Omega)}$, and also define the concentration of each species as the copy number of molecules divided by volume $\Omega$:

$$\bold{x}=\frac{\bold{X}^{(\Omega)}}{\Omega}.$$

The expanded system is subject to the same chemical reactions as the unit-volume system. As for propensity functions, note that within an arbitrary unit volume within the expanded system, the propensity function of the $k$th reaction should roughly equal $a_k (\bold{x})$. If the interactions between different parts of the expanded system are relatively insignificant, then the propensity function of the $k$th reaction in the expanded system of size $\Omega$ should roughly equal $\Omega a_k (\bold{x})$. Based on this argument, we can assume that the propensity function of the expanded system must be of the following form:

\begin{equation}a^{(\Omega)}_k (\bold{X}^{(\Omega)}) =\Omega f_k (\bold{x})+o(\Omega), \label{Expansion_PropFunc}\end{equation}


\noindent  where $a^{(\Omega)}_k (\bold{X}^{(\Omega)})$ can be called the microscopic propensity function and $f_k (\bold{x})$ can be referred as the macroscopic propensity function. 

It can then can be shown that as $\Omega$ goes to $\infty$, the normalized trajectory of the expanded system (or equivalently, the trajectory of the concentration) will converge to the trajectory defined by the following deterministic differential equation (Oppenheim, Shuler and Weiss, 1969; Kurtz, 1970; Kurtz, 1972):

\begin{equation}\frac{d \bold{x}}{d t}=\sum_{k=1}^K f_k(\bold{x}) \boldsymbol{\xi}_k,\label{Deterministic_Limit}\end{equation}

\noindent which is also known as the macroscopic equation. In contrast to the CME where the system state is unit-less, the concentration $\bold{x}$ in macroscopic equation has unit $V^{-1}$. Correspondingly, the unit of macroscopic propensity function $f_k(\bold{x})$ is  $s^{-1}V^{-1}$ and $\boldsymbol{\xi}_K$ is an unit-less vector. 

In the various examples discussed in the last section (including the propensity function in the form of function (\ref{Propensity_Func}) and the nonlinear rational propensity function under QSSA), we have set $\Omega=1$ for the sake of simplicity. If we introduce arbitrary $\Omega$, it is easy to see that the microscopic propensity functions discussed in this example fit the general assumption of equation (\ref{Deterministic_Limit}) with reminder terms $o(\Omega)$ equal to 0. The macroscopic propensity functions in these examples would then be identical in form to the microscopic propensity functions with volume $1$. It is also worth noting that, the rational propensity function (\ref{rational_func}) under QSSA can be regarded as a deterministic approximation. Thus, strictly speaking, the $\bold{X}$ term in (\ref{rational_func}) actually represents the concentration $\bold{x}$ normalized by the volume $\Omega$. \newline

\textbf{Example:} In model \ref{Model_Dogma}, if we use $x_R, x_P$ to denote the concentration of RNA and protein respectively, then under the macroscopic deterministic approximation we have the following differential equations:

\begin{align*}
\frac{dx_R}{dt}=\tau_R-\lambda_Rx_R,\ \  \ 
\frac{dx_P}{dt}=\tau_Px_R-\lambda_Px_P.
\end{align*}

Similarly, in the repressor model \ref{Nonlinear_Repression}, if we use $x_{R_1}$ to represent the mRNA concentration of the first gene, and $x_{P_2}$ to represent the protein concentration of the second gene, the linear approximation will yield:

\begin{align*}
\frac{dx_{R_1}}{dt}=\tau_R \frac{b_0}{1+c_1 x_{P_2}^2}-\lambda_{R_1}x_{R_1}.
\end{align*}
\


This result connects the CME and deterministic differential equation and validates the application of the deterministic model for describing an inherently stochastic system such as a cell. Nonetheless, there are quite a few issues regarding the application of a deterministic approximation. 

First of all, in order for the deterministic limit to apply, the system size, as well as the copy numbers of each molecular species must be sufficiently large so that the fluctuations is relatively insignificant. However, due to the low-copy-number effect at the single-cell level, such assumption might not hold. 

\begin{figure}[htbp]
\begin{center}
\includegraphics[width=1\textwidth]{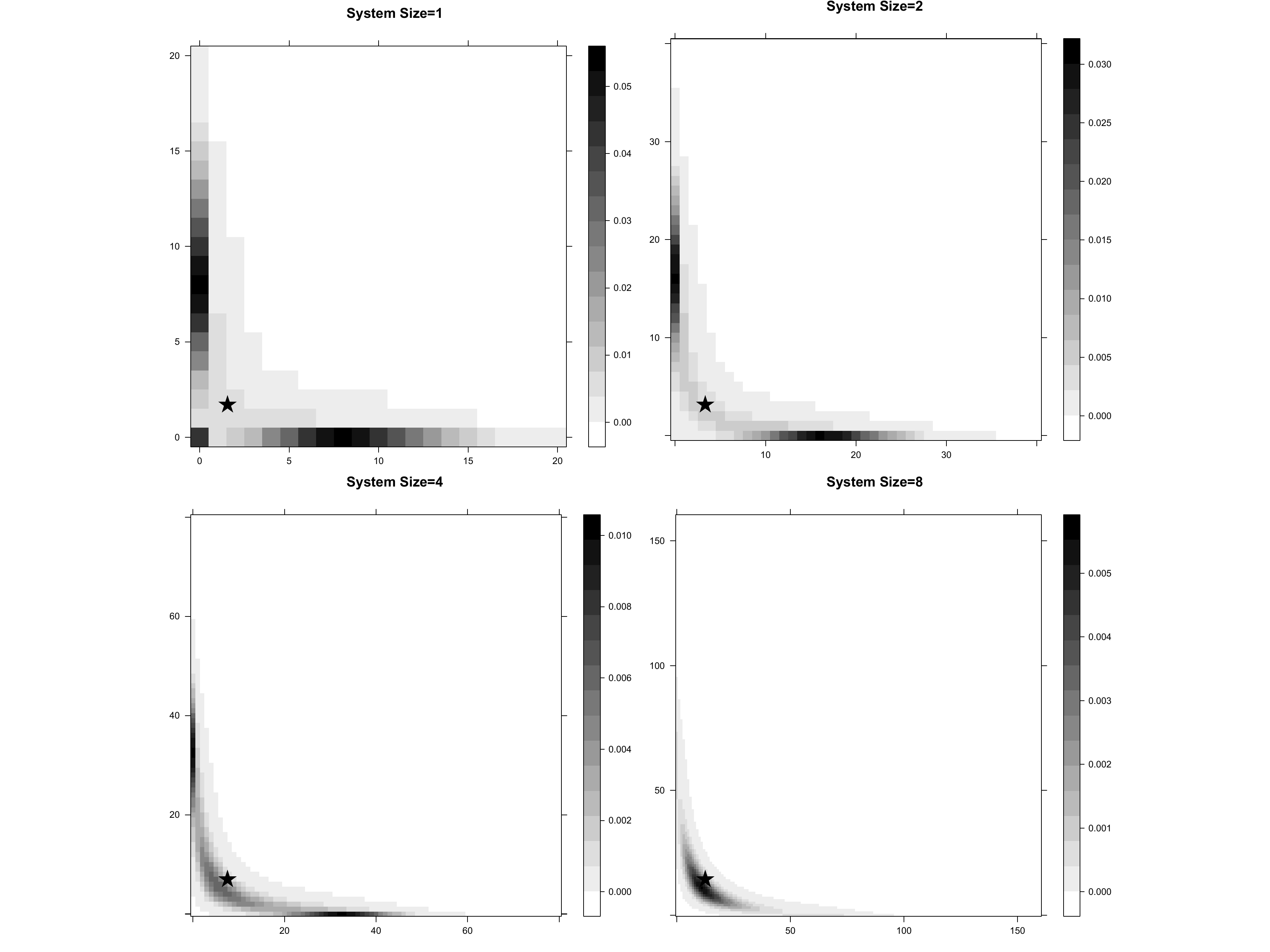}
\caption{Equilibrium Distribution and Size of System}
\label{Dist_Systemsize_figure}
\end{center}
\end{figure}

Second, the qualitative behaviors of a stochastic system and its deterministic limit can be quite different (Qian, Shi, and Xing, 2009; Vellela and Qian, 2009; Bishop and Qian, 2010), and the characteristics of the system might also change dramatically as the system size increases. This is particular true in a CME system with multiple modes. Here are several examples:  1) Certain gene regulatory systems exhibit bimodal behavior, and may stochastically switch between modes as the systems evolve (Ozbudak, \textit{et al.,} 2004; Dubnau and Losick, 2006; Mettetal and van Oudenaarden, 2007; Kuwahara and Soyer 2012). Such switching behavior can be captured by a CME model but cannot be reproduced under a deterministic model. 2) In certain regulatory systems with multiple modes, as the system size increases, the dynamic of the system would shift toward the ``most likely" path. And, at the same time, the chance of taking other trajectories diminishes. Such a phenomenon has been observed in the bacterium \textit{Bacillus subtilis}. This bacterium often remains in a dormant state but can stochastically transit into a ``competent" state and gain the ability to capture DNA from surrounding environment (S\"{u}el \textit{et al.,} 2006). Through introducing a defect in the cell division mechanisms, ``super" cells that consist of multiple individual cells sharing a cytoplasm can be formed. It has been observed that, the probability of transiting into a ``competent" state decreases as the size of the ``super" cell increases (S\"{u}el \textit{et al.,} 2007). 3) As the system size increases, old modes might disappear and new modes might emerge. Reader can refer to the case discussed at the end of this section and Figure \ref{Dist_Systemsize_figure} for a numerical example. 

Third, in practice, a cell only contains a fixed number of genes and it is not expandable. Consequently, while a CME system can be use to represent a cellular system that may exist naturally, its deterministic limit only represents the property of an imaginary system without a physical counterpart. 

In light of these issues with the application of deterministic approximation, caution should be taken when applying a deterministic model for the study of single-cell-level data and when comparing inference results based on ensemble-level data and single-cell-level data. If we treat a single cell as a stochastic system, then the observed single-cell-level data should be regarded as independent realizations of the corresponding stochastic model and the ensemble-level data would represent the average state of the same stochastic model. In contrast, the deterministic limit we discussed above is related to the most likely system state, which can be quite different from the average state, especially when multimodality is present. Thus, the application of a stochastic model in single-cell-level analysis would not only allow us to fully utilize information information contained in the observed noise but may also offer a more realistic understanding of the underlying system. 
\newline

Between the discrete CME model in which noise is treated as intrinsic and the continuous deterministic model that ignores the noise, a compromise is the development of a stochastic approximation over a continuous state space. Depending on how ``stochasticity" is modeled, different approaches, such as linear noise approximation (LNA) and stochastic differential equation (SDE, or the Langevin Equation), can be used. 

We will firstly discuss LNA through the famous system size expansion approach (Van Kampan, 1976, 2007). This approach starts with the following \textit{ansatz}:  the probability function $\P_t(\bold{X}^{(\Omega)})$ has a sharp peak at position of the order $\Omega$, with the width of the order $\Omega^{1/2}$. This \textit{ansatz} will then allow us to represent $\bold{X}^{(\Omega)}$ as:

$$\bold{X}^{(\Omega)}=\Omega \bold{x}(t)+\Omega^{1/2} \bold{y} (t).$$

The first term $\bold{x}(t)$ is the solution of the deterministic equation (\ref{Deterministic_Limit}) and the second term $\bold{y} (t)$ represents stochastic fluctuation whose magnitude often depends on $\bold{x}(t)$. As discussed in Kubo, Matsuo and Kitahara (1973), this \textit{ansatz} essentially assumes that the distribution of system state would retain the uni-modal and bell-shaped characteristics throughout its time course. Under this representation, the trajectory of the system can be decomposed as the main deterministic trajectory determined by (\ref{Deterministic_Limit}) plus a minor stochastic component. The evolution of stochastic component $\bold{y} (t)$ can be solved by noting that, under the \textit{ansatz}, the distribution function $\P_t(\bold{X}^{(\Omega)})$ can be represented as $\boldsymbol{\Pi}_t(\bold{y})$, the distribution function $\bold{y}$. And the time derivative of $\boldsymbol{\Pi}_t(\bold{y})$ follows:

$$\frac{\partial }{\partial t}  \boldsymbol{\Pi}_t(\bold{y})=\frac{\partial }{\partial t} \P_t(\bold{X}^{(\Omega)})+\Omega^{1/2} \sum_{i=1}^M \frac{d x_i}{dt} \frac{\partial \boldsymbol{\Pi}_t}{\partial y_i}.$$

The system size expansion approach will expand the first term on the right-hand side of the above equality based on the order of $\Omega$. In this expansion, the largest terms are of the order $\Omega^{1/2}$, which can be canceled out by the second term on the right hand side of the above equality given that the macroscopic deterministic equation (\ref{Deterministic_Limit}) holds. This fact allows us to focus on the terms of the order $\Omega^{0}$ and discards other terms with lower orders (which would vanish as $\Omega$ approaches infinity). This procedure would then yield the following equality:

$$\frac{\partial }{\partial t}\boldsymbol{\Pi}_t(\bold{y})= -\sum_{i,j=1}^M  A_{ij} \frac{\partial}{\partial y_i}[ y_j \boldsymbol{\Pi}_t]+\frac{1}{2} \sum_{i,j=1}^M B_{ij} \frac{\partial^2}{\partial y_i y_j} \boldsymbol{\Pi}_t,$$

\noindent where 

$$A_{ij}= \sum_{k=1}^K \xi_{ki} \frac{\partial }{\partial x_j} f_k(\bold{x}),\ \ B_{ij}(\bold{x})=\sum_{k=1}^K \xi_{ki} \xi_{kj} f_k(\bold{x}),$$

\noindent and $\xi_{ki}$ represents the $i$th entry of vector $\boldsymbol{\xi}_k $.

%
%
%
%
%

Conditioning on the deterministic trajectory $\bold{x}$, this formula is a linear Fokker-Planck equation whose solution is Gaussian. Thus, under LAN, we can first solve the macroscopic equation to obtain the deterministic trajectory. Then we can construct differential equations on the first and second moments of $\bold{y}$ to solve the mean vector and covariance matrix (Van Kampan, 2007). Such strategy would then allow us to determine the distribution of system at any time $t$ as well as the equilibrium distribution. 

Using the Ito interpretation, the above Fokker-Planck equation is also equivalent to the following Stochastic differential equation (Komorowski,\textit{et al.,} 2009):

$$d \bold{y}_t= \bold{A} \bold{y}  dt+ \boldsymbol{\Sigma} d\bold{W}_t $$

\noindent where $\bold{A}$ is a $M\times M$ square matrix whose entry at $i$th row and $j$th column is $A_{ij}$,  $\bold{W}_t$ is a $K$ dimensional Wiener process, and $\boldsymbol{\Sigma}$ is a $M \times K$ dimensional matrix whose entry at $i$th row and $k$th column is $\xi_{ki} \sqrt{ f_k(\bold{x})} $. Both drift and diffusion coefficients only depend on $t$ through $\bold{x}$. 

%
%
%
\

\textbf{Example:}  In Model \ref{Model_Dogma}, the LAN approximation of the noise term $\bold{y}=(y_R, y_P)^T$ can be described by the following equation:

\begin{align*}
\frac{\partial }{\partial t}\boldsymbol{\Pi}_t(\bold{y}) & = \frac{\partial}{\partial y_R} [\lambda_R y_R \boldsymbol{\Pi}_t ]- \frac{\partial}{\partial y_P} [( \tau_Py_R-\lambda_P y_P) \boldsymbol{\Pi}_t ] \\ & +\frac{1}{2}  (\tau_R+\lambda_R x_R) \frac{\partial^2}{\partial y_R^2} \boldsymbol{\Pi}_t
+\frac{1}{2} (\tau_Px_R+\lambda_Px_P) \frac{\partial^2}{\partial y_P^2} \boldsymbol{\Pi}_t,
\end{align*}

\noindent and the equivalent stochastic differential equation are:

\begin{align*}
dy_R & = -\lambda_R y_R dt+ \sqrt{\tau_R} dW_1-\sqrt{\lambda_Px_R} dW_3,\\
dy_P & = (\tau_Py_R-\lambda_P y_P) dt+ \sqrt{\tau_Px_R} dW_2-\sqrt{\lambda_Px_P} dW_4.\\
\end{align*}
\
It is worth noting that although the fluctuation in LNA appears as an additive term to the deterministic component, the strength of the fluctuation depends on the deterministic trajectory and thus reflects the underlying mechanisms of the CME system. In this regard, the fluctuation term in LNA should be treated as multiplicity noise rather than additive noise. In fact, as demonstrated in the work of Frigola, \textit{et al.,} (2012), additive noise is often insufficient to appropriately model the underlying complex system. Nonetheless, as the center of the approximated distribution must follow the deterministic trajectory, the effectiveness of LNA depends on how well the deterministic approximation works. In particular, LNA would fail to capture the key characteristics of the system if the system contains multiple steady states or moves past a critical point (Baras, Mansour, and Pearson, 1996). 
\newline

Another continuous approximation is based on the so-called Kramers-Moyal expansion, a Taylor series expansion of the CME developed by Kramers (1940) and Moyal (1949). In this approach, the discrete CME is approximated with a continuous random process whose evolution can be described with a nonlinear Fokker-Planck equation (Gardiner, 2004). This approximation scheme can be equivalently represented by the following stochastic differential equation (Gillespie, 2000):

$$d \bold{X}_t= \sum_{k=1}^K \boldsymbol{\xi}_k a_k(\bold{X_t})  dt+ (\boldsymbol{\xi}_1 \sqrt{a_1(\bold{X_t})}, \boldsymbol{\xi}_2 \sqrt{a_2(\bold{X_t})},\cdots, \boldsymbol{\xi}_K \sqrt{a_K(\bold{X_t})} ) d\bold{W}_t $$

\noindent where $\bold{W}_t$ is a $K$ dimensional Wiener process. In the following discussion, we will use SDE to refer this approximation.\newline

\textbf{Example:} The SDE approximation of Model \ref{Model_Dogma} is:

\begin{align*}
dX_R & = (\tau_R-\lambda_R X_R) dt+ \sqrt{\tau_R}dW_1-\sqrt{\lambda_RX_R} dW_3,\\
dX_P & = (\tau_PX_R-\lambda_P X_P) dt+ \sqrt{\tau_PX_R}dW_2-\sqrt{\lambda_PX_P} dW_4.\\
\end{align*}
\
An intuitive explanation of SDE approximation can be obtained from equation (\ref{PoissonRep}), the Poisson process presentation of CME. Between $(t,t+\Delta t)$, the number of firing of the $k$th reactions is $Y_k [\int_{t}^{t+\Delta t} a_k(\bold{X} (s)) ds]$, which can be approximated with Poisson distributed random variable with rate $\Delta t a_k(\bold{X} (t))$. If we assume that the system size is sufficiently large, $\Delta t a_k(\bold{X} (t))$ would also be large enough to guarantee a Normal approximation. Rigorous discussion on the accuracy of SDE approximation can be found in the work of Kurtz (1978) and Grima, Thomas and Straube (2011).

\begin{figure}[htbp]
\begin{center}
\includegraphics[width=1\textwidth]{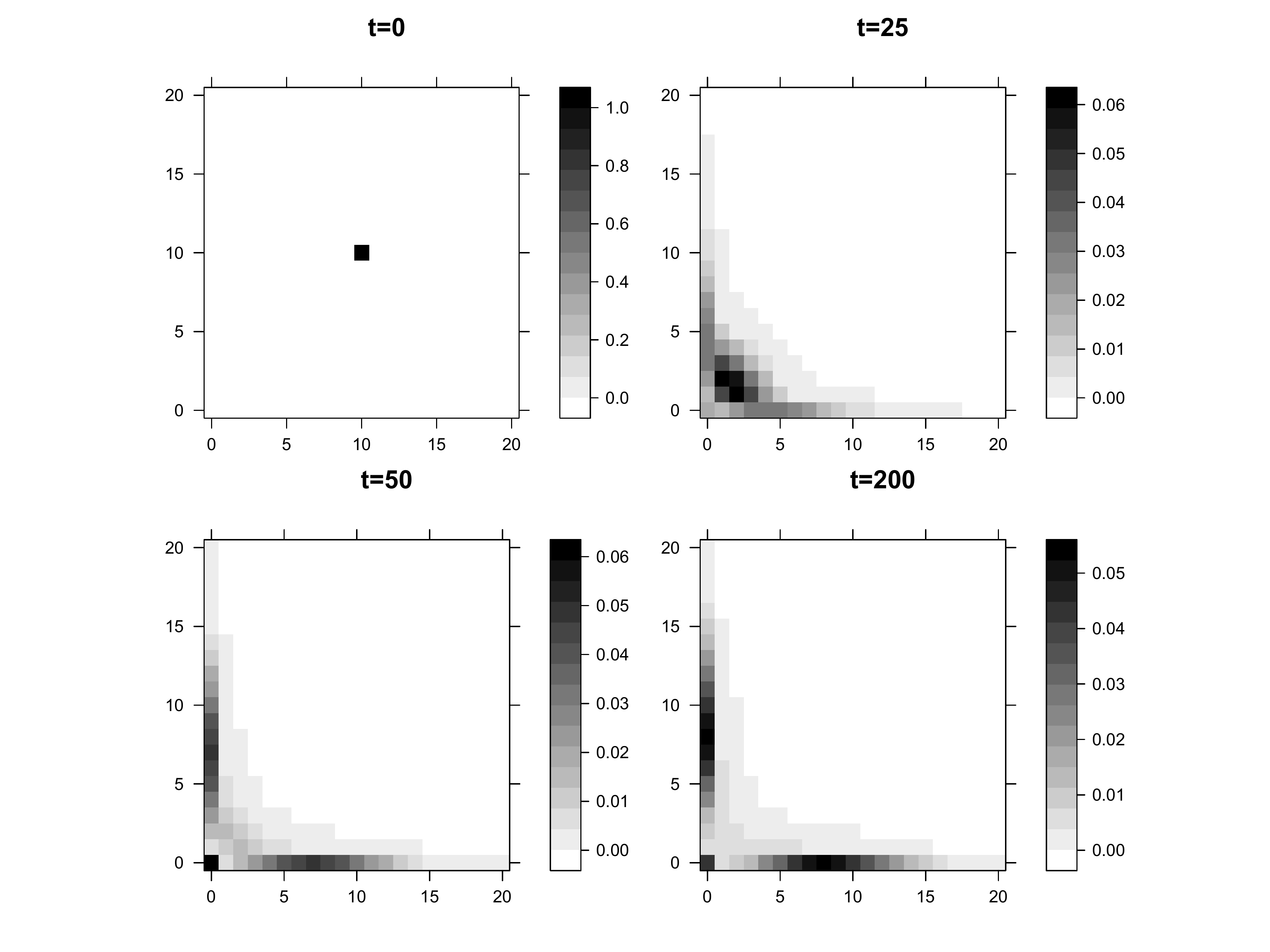}
\caption{Dynamic Evolution of a CME system}
\label{Dynamic_CME_figure}
\end{center}
\end{figure}

\begin{figure}[htbp]
\begin{center}
\includegraphics[width=0.9\textwidth]{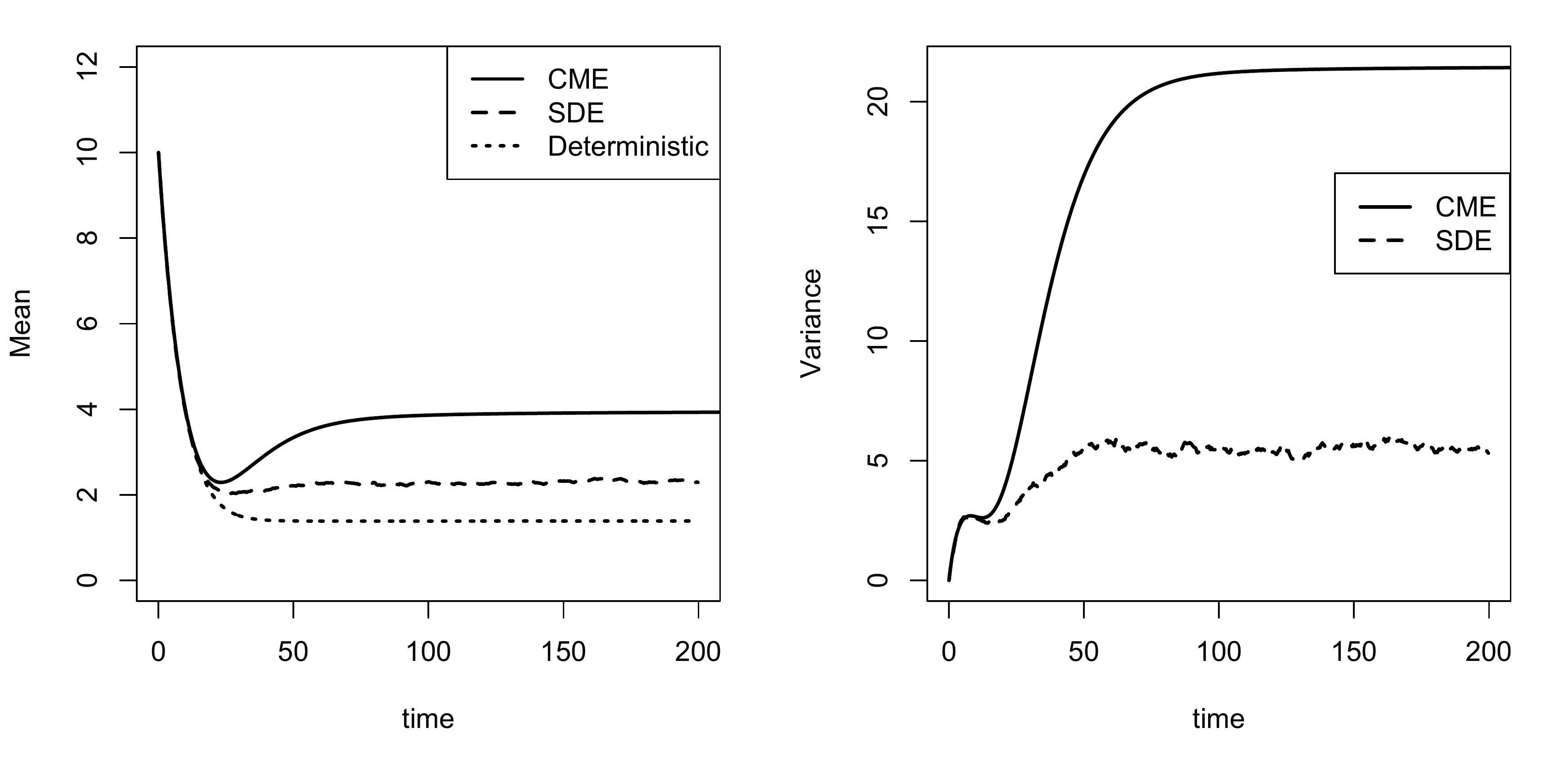}
\caption{Evolution of Mean and Variance}
\label{mean_var_figure}
\end{center}
\end{figure}

The key advantage of SDE over LNA is that it does not require that the stochastic process evolve along the trajectory determined by the macroscopic equation; thus it has the potential to approximate the CME better when the underlying distribution is significantly away from a unimodal distribution. For instance, in certain systems such as the genetic toggle switch (Gardner, Cantor and Collins, 2000), even if the initial distribution of the system is unimodal, it may evolve into a multimodal distribution as the system passes through a critical point. However, the coupled deterministic system would only evolve towards one of the modes determined by the initial condition. Correspondingly, the LNA may not capture the qualitative behavior of the original system well. In contrast, the diffusive nature of SDE would still allow the probability mass to be distributed towards different modes; therefore, it may still capture the characteristics of the original system. Still, the effectiveness of SDE approximation is limited when the discreteness of the CME cannot be easily ignored. The potential discrepancy between CME and SDE approximation has been explored in a simulation study (Baras, Mansour and Perason, 1996). Moreover, in an experimental study on a genetic toggle switch (Ma, \textit{et al.,} 2012), it has also been observed that the system can enter a third stable state in which both gene express at a very low level. Such a phenomenon cannot be predicted by the SDE model but can be explained using the CME model. Readers can also refer to the case discussed below as well as Figure \ref{Dynamic_CME_figure} and Figure \ref{mean_var_figure} for a numerical example. \newline

Before we conclude this section, we would like to present a numerical example focusing on the discrepancy between CME, deterministic approximation and SDE approximation. This example is chosen to emphasize the potential impact of the discreetness and multimodality on the approximation of CME system.

\textbf{Example:} Here we consider a two-gene version of Model \ref{Simplified_Rational_Model}. In this example, we assume that the system size $\Omega=1$, and set $\tau_1=\tau_2=1$, $\lambda_1=\lambda_2=0.1$. The rational propensity functions of two genes are assumed to be of the following form:

$$F_1(X_1,X_2)=\frac{0.01+X_1}{1+X_1+4X_1X_2}$$
$$F_2(X_1,X_2)=\frac{0.01+X_2}{1+X_2+4X_1X_2}.$$

In this system, both genes serve as their own activators. However, the complex formed by protein molecules from each gene would serve as the repressor to both genes. 

In Figure \ref{Dist_Systemsize_figure}, we draw the equilibrium distribution of this CME system, as well as the equilibrium distributions when the system size is increased to $2,4,$ and $8$, respectively. The stable point of the corresponding deterministic equation is also plotted for reference (as a star).  As we can see from this plot, although the four different CME systems are all approximated by the same deterministic equation, they exhibit very different characteristics.  When the system size is 1, there are three modes in the equilibrium distributions, which corresponds to the cases where both genes are inactive and one of the two genes is active. The stable point of the deterministic approximation, however, is located in a valley between these three modes. Furthermore, the following phenomena can be observed as the system size increases: 1) the mode where both genes are inactive disappears; 2) the probability mass around the other two modes starts to decrease; and 3) a new mode starts to form near the stable point of the deterministic equation. 

In Figure \ref{Dynamic_CME_figure}, we plot the dynamic evolution of this CME system between $t=0$ and $t=200$ (at this point the system has effectively reach the equilibrium state). The initial distribution is set as a point mass at $(10,10)$. Roughly speaking, at the beginning stage of the evolution (before $t=25$), the center of the distribution keeps moving towards to the lower-left corner of the state space and the shape of distribution remain unimodal. However, as the system reaches the saddle point as shown in the sub-plot at $t=25$, the distribution starts to differentiate into three modes. 

In Figure \ref{mean_var_figure}, we compare the evolution of this CME system with its deterministic and SDE approximations by plotting the evolution of the mean and variance of component $X_1$ $t=0$ and $t=200$. As seen in this figure, deterministic and SDE approximations work reasonable well as long as the distribution of CME remains unimodal. In contrast, as the system moves to the lower-left corner of the state space (after $t=25$) where the discreteness and multimodality starts to play major roles, both deterministic and SDE approximations fail to predict the evolution of original CME system. The SDE approximation still performs better than deterministic approximation in term of the mean, though. \newline

In summary, while an unmodified CME model can capture the discreteness, nonlinearity and intrinsic stochasticity of a biological system, the different approximation schemes discussed above would either ignore or compromise in terms of modeling these traits. In this regard, an unmodified CME model would be more closer to the ``truth" than other approximation schemes constructed under the same principles. Still, the discussion of the comparison between the CME model and its approximations is still quite limited in the current literatures. We hope our discussion stimulates research interests in this topic. 

\section{Simulation of the Chemical Master Equation}

Simulation is an indispensable tool for studying the properties of a CME system as the analytical solution of a CME is usually impossible to obtain. In particular, the distribution of the system state at any given time or at equilibrium can be reconstructed with samples collected from many independent trajectories. Utilizing the fact that a CME system is driven by the successive firings of different reactions and the waiting times follow exponential distributions, Gillespie (1976,1977) proposed the following exact algorithm for simulating the realizations of trajectories of a CME system:

\begin{itemize} \itemsep0pt
\item{Step 1}: Set the starting system state as $\bold{X}(0)$, and set time $t=0;$
\item{Step 2}: Calculate the propensity functions $a_k(\bold{X}(t))$ for $k=1, 2,\cdots, K;$
\item{Step 3}: For each $k$, generate an exponentially distributed random variable $\tau_k$ with rate $a_k(\bold{X}(t))$, and set $\tau=\min(\tau_1,\cdots, \tau_K)$.
\item{Step 4}: Set $I=\underset{k}{\text{argmin}} \ \tau_k$. 
\item{Step 5}: Advance the system by updating $\bold{X}\rightarrow \bold{X}+\xi_I$, and $t\rightarrow t+\tau$;
\item{Step 5}: Stop if preset conditions are met. Otherwise, return to step 2. 
\end{itemize}

This algorithm is called the direct method. In particular, steps 3 and 4 are used to determine when and which of the the $K$ reactions will fire next. In this algorithm,  $K$ uniform distributed random variables need to be generated in each iteration. This cost can be reduced significantly by utilizing the memoryless property of the exponential distribution. That is, 
the waiting time for the next reaction to fire follows a exponential distribution with rate $\sum_k a_k (\bold{X(t)})$, and the chance that the $k$th reaction is the next reaction is proportional to $ a_k (\bold{X(t)})$. Then, steps 3-4 of the direct method can be improved by the following procedure (Gillespie, 1976,1977):

\begin{itemize} \itemsep0pt
\item{Step 3*}: Generate an exponentially distributed random variable $\tau$ with rate $\sum_k a_k (\bold{X(t)})$ and a uniform random variable $U$.
\item{Step 4*}: Let $I=k$ so that 
$$\frac{\sum_{j=1}^{k-1} a_j (\bold{X(t)}) }{\sum_{k=1}^K a_k (\bold{X(t)})}\leq U <\frac{\sum_{j=1}^{k} a_j (\bold{X(t)}) }{\sum_{k=1}^K a_k (\bold{X(t)})},$$
\noindent where $\sum_{j=1}^{k-1} a_j (\bold{X(t)})=0$ when $k=1$. 
\end{itemize}

The revised algorithm is firstly named the first reaction method, but it is commonly known as the stochastic simulation algorithm (SSA) or Gillespie's algorithm.  In each iteration of this algorithm, we only need to generate two uniform random variables. However, the computational cost related to the nonrandom elements, including updating all propensity functions in step 2 and selecting the first reaction in step 4*, is still of the order $K$ for each iteration. 

The performance of SSA can be improved using the next reaction method (Gibson and Bruck, 2000) through two major innovations. First, by exploiting the memoryless property, we can focus on the absolute time frame starting from the beginning rather than the relative waiting times between successive reactions. Such exploitation can reduce the number of uniform random variables needed for each iteration from 2 to 1 (with the exception of the first iteration). Second, by storing all relevant information in a tree structure, we can minimize the computing cost in determining the next reaction and avoid unnecessary reevaluation of the propensity functions.  This strategy can reduce the computational complexity related the nonrandom elements to the order of $\log(K)$. Anderson (2007) reformulated the next reaction method based on the Poisson process presentation of CME and suggested that such an approach can be used to simulate a nonMarkov system, such as a system with delayed reaction time or a system with time-dependent propensity functions. Additional improvement can also be achieved by implementing a more sophisticated data structure reflecting the topology of different reactions, such as the LOLCAT method (Indurkhya and Beal, 2010). 
 \newline

While the aforementioned algorithms can be used to efficiently simulate exact trajectories of a CME, the computational cost of simulating a single trajectory can still be very high as every firing of reactions during the evolution of the system is taken into account. Considering the fact that the distribution of the system state can only be reasonably reconstructed based on a large number of independent trajectories, the exact methods may be unrealistic and approximation schemes are often needed.

One common strategy is to assume that the propensity functions remain constant during a certain period $(t, t+\tau)$ (for instance, when $\bold{X} (t)$ is large compared to all $\boldsymbol{\xi}_k$, the change due to the firings would be insignificant), then the number of firing the $k$th reaction during this period would follow a Poisson distribution (see the Poisson representation in equation  (\ref{PoissonRep})).  Then we may update the system state from $t$ to $t+\tau$ directly through the following approximation: 

$$\bold{X} (t+\tau)\approx \bold{X} (t)+ \sum_{k=1}^K \boldsymbol{\xi}_k M_k,$$

\noindent where $M_k$ is independent Poisson distributed random variable with rate $\tau a_k(\bold{X}(t) )$.

This method is called $\tau$-leaping algorithm (Gillespie, 2001). Generally speaking, a large $\tau$ means a fast computation time but low precision. Consequently, the performance of this algorithm depends on the appropriate choice of step size $\tau$, which should be updated dynamically during the course of simulation (Gillespie, 2001; Gillespie and Petzold, 2003; Cao, Gillespie and Petzold, 2006). Moreover, the effectiveness of this algorithm can also be improved by using $a_k(\bold{X}(t^*))$, where $t^*$ is chosen as a ``mid-point" between $t$ and $t+\tau$, rather than $a_k(\bold{X}(t))$ to approximate the propensity function during $(t, t+\tau)$ (Gillespie, 2001).  This $\tau$-leaping algorithm can also be combined with the next reaction method for greater efficiency (Puchalka and Kiezek, 2004). Another similar method is known as the R-leaping method (Auger, Chatelain and Koumoutsakos, 2006), in which the algorithm is set to leap forward by a fixed number of firings rather than by a fixed time period. Under the similar assumption on propensity function, the elapsed time for making R-leaps follows Gamma distribution. 

Another strategy utilizes the QSSA. In many applications, the propensity functions often have vastly different timescales so certain reactions fire much more frequently than others.  Not surprisingly, in an exact simulation, most of the computing efforts will be invested in the ``fast" reactions, while the evolution of the system is often determined by the ``slow" reactions. In this regard, computational efficiency can be improved by using an exact algorithm for the ``slow" reactions but applying an approximated scheme to handle the ``fast" reactions. 

For instance, in the work of Haseltine and Rowling (2002), the evolution of the system under ``fast" reactions is approximated using the stochastic differential equation. In the maximum time step method (Puchalka and Kiezek 2004), $\tau$-leaping algorithm is used to handle ``fast" reactions. This method categorizes reactions into ``fast" and ``slow" groups based on two major criteria: first, the copy numbers of species involving ``fast" reactions must be greater than a threshold; second, the chance of a ``fast" reaction firing first must be greater than a given threshold. Given that these conditions change as a system evolves, the classification of  ``fast" and ``slow" reactions is dynamically updated throughout the course of simulation. In the slow-scale stochastic simulation algorithm (Cao, Gillespie, Petzold, 2005), not only are the reactions allocated into two categories, but the species involved in CME system are also classified. ``Fast" species are defined as species impacted by one or several ``fast" reactions; the rest are ``slow" species. QSSA can then be used to determine the conditional distribution of  ``fast" species given the current state of ``slow" species. And the evolution of ``slow" species can be simulated based on the reduced CME  (\ref{QSSA_discrete}).\newline
%
%

The above discussion has focused on how to simulate independent trajectories of a CME. If the goal is to estimate the probability of a particular event (such as the probability that the copy number of a given species reaches a threshold within a given period), the technique of important sampling can be used to improve the naive estimator. The weighted SSA is the first algorithm that incorporates important sampling into SSA (Kuwaharaa and Mura, 2008). In this algorithm, the propensity functions in step 4* of SSA are scaled by predetermined constants, changing the relative priorities of reactions. Appropriate scale constants can dramatically increase the chance of occurrence of the event of interests. The bias introduced in this process is adjusted by reweighting the simulated trajectory accordingly. Within such a framework, the confidence intervals of the estimator can also be estimated (Gillespie,  Roh, and Petzold, 2009). A few modifications of the weighted SSA were developed, mainly focusing on improving the choice of scale constants. The state-dependent SSA (Roh, Gillespie, and Petzold 2010) allows the scale constants to be dynamically updated during the course of simulation. In the double-weighted SSA (Daigle, \textit{et al.,} 2011), scale constants are also applied in step 3* of SSA to modify the distributions of of waiting times between firings. The purpose of this strategy is to allow cross-entropy method to be applied to determine the optimal choice of scale constants. The two strategies discussed above are implemented together in the state-dependent doubly weighted SSA to achieve greater efficiency (Roh, \textit{et al.,} 2011).

For more discussion on the issue of simulating CME system, readers can also refer to the following more in-depth review work: Gillespie (2007), Gillespie, Hollander and Petzold (2013).

\section{Solving the Chemical Master Equation by Analytical and Numerical Methods}

In order to better understand a CME system, to infer the unknown parameters in a CME model from observed data or to compare competing models with different level of details, it is often necessary to establish a quantitive relationship between the model parameters and the distribution of the system state. For such a purpose, in addition to our discussion of the simulation of a CME system, we will explore how the CME can be solved through analytical and numerical means in this section. In particular, we will focus our discussion on how to obtain the solution to the moment function and the distribution function of the CME. 

\subsection{Moment Function of the Chemical Master Equation}

Considering the fact that the CME is defined on a discrete sample space, it is often easier to analyze moment equations than study the full distribution function. For instance, the moment equality  (\ref{moment}) can be used to establish differential equations of the expectation of any function at time $t$ or at the equilibrium state. \newline

\textbf{Example}: In model \ref{Model_Dogma}, if we apply equation (\ref{moment}) for $H(\bold{X})=X_R^nX_P^m$ where $n,m$ are non-negative integers, we have:
\begin{align*}
& \frac{d}{dt} \E(X_R^nX_P^m)  \\
=& \E[ ((X_R+1)^n-X_R^n)X_P^m \tau_R] +\E[((X_R-1)^n-X_R^n) X_P^m \lambda_R X_R] \\
+& \E[X_R^n ((X_P+1)^m-X_P^m) \tau_P X_R]+ \E[X_R^n ((X_P-1)^m-X_P^m) \lambda_P X_P].
\end{align*}

As the orders of the moments in the right-hand of the differential equations are equal or lower than the order of moments in the derivative terms, the general solution of $\E(X_R^nX_P^m)$ can be found by an iterative procedure. We start with $\E(X_R^n)$. If we set $m=0$, the above equation can be reduced to only include the moments of $X_R$ up to the $n$th order. If we further set $n=1$, we can solve $\E(X_R)$. This result can be further used to solve $\E(X_R^2)$ when we let $n=2$. Higher moments of $\E(X_R^n)$ can be found in similar fashion. Similarly, $\E(X_P)$ can be found based on  the solution of $\E(X_R)$, and $\E(X_RX_P)$ can be found using $\E(X_P), \E(X_R)$ and $\E(X_R^2)$... The solutions will be analytical functions of the rate parameters and can be used for inference purpose \newline

Unfortunately, such an approach is only possible when all propensity functions are either constant or linear functions. The issue of solving moment functions is much more complicated with nonlinear propensity functions.

First of all, let us consider the stoichiometric network in which all the propensity functions are of the form (\ref{Propensity_Func}). Equation (\ref{moment}) can still be applied for establishing differential equations, in which the derivatives of moment are represented as the linear functions of other moments. Nonetheless, as long as some propensity functions are nonlinear, the orders of certain terms in the right-hand of the equality will be higher than the order of moments in the derivative term. In short, let us use $\boldsymbol{\mu}_n$ to represents the vector of moments of $\boldsymbol{X}$ up to order $n$; the ordinary differential equation for $\boldsymbol{\mu}_n$ can be expressed in the following general form:



\begin{equation}\frac{d \boldsymbol{\mu}_n}{d t}=C_n\boldsymbol{\mu}_n+C_n^*\boldsymbol{\mu}_n^*,\label{MMStoModel}\end{equation}

\noindent where $\boldsymbol{\mu}_n^*$ is a vector with moments whose orders are higher than $n$, $C_n$ and $C_n^*$ are matrices whose elements depend on the rate parameters and structure of the propensity functions. More details regarding such method can be found in the work of Engblom (2006), Gillespie (2009), as well as Sotiropoulos and Kaznessis (2011). Due to the presence of higher-order moment terms, this equation can not be solved analytically nor numerically. Approximation approaches collectively known as moment closure method are often used to resolve this issue. In a typical moment closure scheme, higher-order moments $\boldsymbol{\mu}_n^*$ are approximated as functions of lower-order moments $ \boldsymbol{\mu}_n$. Then equation (\ref{MMStoModel}) only includes lower-order moments $\boldsymbol{\mu}_n$ only and is effectively ``closed".  \newline

\textbf{Example}: In Model \ref{Model_Dogma}, if we assume that the protein molecule can form a dimmer, and add the following reaction:

 $$2P  \xrightleftharpoons[k_{-1}]{\,k_{1}\,}  P\text{-}P. $$
 
 Then the moment equation on the free protein molecule $P$ will be:
  \begin{align*}
& \frac{d}{dt} \E(X_P^m)  \\
=& \E[((X_P+1)^m-X_P^m) \tau_P X_R]+ \E[((X_P-1)^m-X_P^m) \lambda_P X_P] \\
+& \E[((X_P+2)^m-X_P^m) k_{-1} X_{P\text{-}P}]+ \E[ ((X_P-2)^m-X_P^m) k_1 X^2_P].
\end{align*}
 
In this equality, while the left-hand-side is the time derivative of the $m$th moment of $X_P$, the right-hand-side involves the $(m+1)$th moment of $X_P$ due to the quadratic propensity function of the newly introduced reaction. Thus, this equality can not be solved by the approach used in the previous example, and approximation must be made to represent $\E(X_P^{m+1})$ as functions of lower moments. \newline

Numerous methods have been developed to define an appropriate closure scheme. The earliest method utilized the relationship between the cumulants and moments. If we ``truncate" all the higher-order cumulants by setting them to 0, we can obtain linear equalities of moments and express higher-order moments as linear combinations of lower-order moments. This method was first introduced by Whittle (1957): all the third- and higher-order cumulants were set to 0. Similar approach was also discussed in the work of Matis and Kiffe (1996) and N\aa sell (2003b). Alternatively, we may also expand the propensity function around the mean, and establish a closure scheme by truncating the higher-order terms in the Taylor expansion (Lee, Kim and Kim, 2009; Ale, Kirk, and Stumpf, 2013; Lee, 2013). In these approaches, the truncation is effectively made based on the central moments rather than cumulants. The accuracy of the truncation-based approximation methods were discussed in Lee, Kim and Kim (2009) and Grima (2012).

Since truncating all the third- and higher-order cumulants is equivalent to assuming that the solution of CME follows a Normal distribution, we may also define a moment closure scheme by imposing other distributional assumptions on the solution of the CME. For single-variable model, popular distribution choices include Poisson, Log-Normal and Beta-Binomial (N\aa sell, 2003a; Krishnarajah,\textit{et al}., 2005). The latter two distributions are more suitable for modeling a system with a skewed distribution. For multivariable model, mixture distribution (Krishnarajah, \textit{et al.,} 2007), as well as the multivariate version of Normal, Log-Normal and Gamma distribution (Lakators, \textit{et al.,} 2015) can be used. However, such moment closure schemes all suffer a common drawback: the degree of approximation is totally determined by the choice of distribution.

More flexible moment closure schemes in which the degree of approximation is tunable can be found in literature. In the separable derivative-matching moment closure approach (Singh and Hespanha, 2006; Singh and Hespanha 2011), higher-order moments are assumed to be the products of the powers of lower-order moments. The constant exponents used in this representation is determined through matching the time derivatives of the true moment functions with the derivative of approximated moments. If all the third- and higher-order moments are approximated, this procedure would be equivalent to the moment closure method with Log-Normal distributional assumption. In the zero-information moment closure scheme (Smadbeck and Kaznessis, 2013), the solution of the CME is approximated by a distribution satisfying the following conditions: 1) the lower-order moments in this distribution match with the lower-order moments in CME; 2) given condition 1), the entropy of this distribution is maximized. This maximum entropy distribution is then used to approximate the higher-order moments as functions of lower-order moments. \newline

The effectiveness of the aforementioned moment closure methods depends on how well the characteristics of the CME system can be described by lower-order moments. However, when the distribution of the CME system exhibit multimodality, it is often hard to describe the multiple modes, as well as the transitions between modes with lower-order moments alone. Ruess \textit{et al.,} (2011) utilized on the samples generated by stochastic simulation algorithm to approximate the more informative higher-moments. Thomas, Popovi\'{c} and Grima (2014) extended the LNA and showed that the distributions of many CME systems with multimodality can be approximated by mixture of Normal distributions. Their work provides means to construct the moment equalities around individual modes as well as estimate the transition probabilities between modes. 

This limitation could be overcome by using simulation-based approach. In the work of Ruess \textit{et al.,} (2011), the higher-order moments are approximated from samples generated by stochastic simulation algorithm. An extended Kalman filter is also used to increase the efficiency. 
\newline

%

When the propensity functions in the CME system are rational functions, the problem of creating moment equality can be more complicated. Milner, Gillespie and Wilkinson(2011) provided a general solution to such a problem. Roughly speaking, we can firstly multiple both sides of the CME (\ref{CME}) by the common product of denominators of the propensity functions to eliminated the ratios in the equality, so further moment equality can be constructed. \newline

\textbf{Example}: Let us consider a single-gene version of Model \ref{Simplified_Rational_Model}. Suppose that the production function is of the form $F(X)=\frac{b}{1+cX}$, then the corresponding CME can be expressed as:
  \begin{align*}
 \frac{d}{dt} \P(X)  = & -[\tau \frac{b}{1+cX}+\lambda X] \P(X) \\
+& \lambda (X+1) \P(X+1)+\tau \frac{b}{1+c(X-1)} \P(X-1).
\end{align*}
Multiplying both side with $(1+cX)(1+c(X-1))$ and summing up over all values of $X$, we can obtain the following moment equality. 
  \begin{equation*}
(2c-c^2) \frac{d}{dt} \E(X)+c^2 \frac{d}{dt} \E(X^2) = 2\tau bc-\lambda c(1-c)\E(X)+\lambda c^2 \E(X^2) 
\end{equation*}
Equalities involving higher-order moments can be constructed by multiplying $e^{\theta X} (1+cX)(1+c(X-1))$ to the both side of the CME, expanding the exponential terms and collecting terms based on the polynomial order of $\theta$. \newline

In the setting of gene regulation systems with rational rate functions, Achimescu and Lipan (2006) and Raffard, \textit{et al.,} (2008) studied a single gene system with mRNA and protein species similar to Model \ref{Nonlinear_Repression}. After multiplying the CME with the denominator terms of the propensity functions, z-transformation is used to transform the distribution $\P_t(\boldsymbol{X})$ into $F(\boldsymbol{z},t)=\sum_{\boldsymbol{X}} z_i^{X_i} \P_t(\boldsymbol{X})$, whose taylor expansion coefficients are the factorial cumulants. Moment closure schemes can then be established by truncating the higher-order factorial cumulants. Their work can also be applied to CME system driving by an external signal. Chao and Wong (draft paper) studied gene regulatory network consisting of multiple genes as in Model \ref{Simplified_Rational_Model} and shown that exact moment equalities can be constructed for single-gene system. For multiple-gene systems, this result can be used to construct approximated moment equalities that prioritize the precision of marginal moments over the precision of cross moments.\newline


%
%
%

\subsection{Approximating the Distribution Function of Chemical Master Equation}

Although the moment functions can grant us valuable insight into the characteristics of the CME system, complete information can only be gained by investigating the distribution function directly. Many statistical inference procedures also require likelihood functions that are analytically or numerically trackable. In the following discussion, we will review the available methods for estimating the probability distribution of the CME system, including the distribution at an arbitrary time $t$ and the equilibrium distribution.  

The CME is a special case of continuous time Markov Chain. If we could enumerate all states in the state space as a vector $\bold{S}$, and denote the corresponding transition matrix as $\bold{Q}$,  then given initial distribution $\P_0(\boldsymbol{S})$, the distribution of system at time $t$ is

\begin{equation}\P_t(\boldsymbol{S})=e^{\bold{Q}t}\P_0(\boldsymbol{S}).\label{ECME}\end{equation}

\noindent The steady-state distribution can also be calculated: it is the normalized left eigenvector corresponding to the eigenvalue 0 of matrix $\bold{Q}$. 

Nonetheless, due to the CME's discrete nature, a direct evaluation of equation (\ref{ECME}) represents is a tremendous, if not impossible, task, even for a simple CME. In an open system in which the state variables can take any nonnegative integer values, the state space consists of an infinite number of states and the transition rate matrix is of infinite dimension. Even if we impose an upper bound on the value of the state variable, the total number of states can still be astronomical. In contrast, practically speaking, it is also highly unlikely for a biochemical system to visit the whole state space during a limited time period. Similarly, the distribution of a system at any finite time $t$ or the equilibrium distribution is likely concentrated over a small subset of the whole state space. For instance, in Model \ref{Model_Dogma}, even if we limit the copy number of mRNA and protein to less then 100, the full state space would still contain 10000 different states. However, at equilibrium state, the probability mass of the distribution concentrates over a small region around the mode, and the probabilities of other states are practically 0. Even if we wished to study the evolution of the system over time, regardless of the initial condition, this system would evolve quickly toward the mode of equilibrium distribution and have a near-0 probability of ever visiting most of the 10000 states. In this sense, the computational cost would be reduced if we could focus on the states that actually matter.

This idea forms the basis of the finite state projection method (FSP, by Munsky and Khammashb, 2006). In this method, the complete state space $\boldsymbol{S}$ is decomposed to a reduced and finite subspace $\boldsymbol{J}$, termed the "finite projection space", which consists of the states frequently visited by the system, and its complement. The transition matrix $\bold{Q}$ can then be partitioned accordingly. If we denote the resulting block matrix corresponding to $\boldsymbol{J}$ as $\bold{Q}_J$, and assume that $\boldsymbol{J}$ contains the support of the initial distribution, then the distribution of the system over the projection space at time $t$ can be approximated as:

\begin{equation}\P^*_t(\boldsymbol{J})=e^{\bold{Q}_J}\P_0(\boldsymbol{J}),\label{FSP}\end{equation}

\noindent where $\P_0(\boldsymbol{J})$ the partitioned vector of initial distribution $\P_0(\boldsymbol{S})$ based on $\boldsymbol{J}$.

The accuracy of approximation will increase monotonically as additional states are added into $\boldsymbol{J}$. And the quality of FSP approximation can be evaluated using the following result. Let $\P_t(\boldsymbol{J})$ be the partitioned vector of the exact distribution $\P_t(\boldsymbol{S})$ at time $t$, then for any $\varepsilon>0,$

\begin{equation} \text{if }  \bold{1}^T\P^*_t(\boldsymbol{J}) \geq 1-\varepsilon, \text{ then } \bold{0}  \leq \P_t(\boldsymbol{J})- \P^*_t(\boldsymbol{J}) \leq \varepsilon \bold{1}\label{FSP_Error}.\end{equation}

Munsky and Khammashb (2006) thus suggested an iterative procedure: starting from a relatively small subset, the finite projection space is iteratively expanded by adding new states based on the reachability from the current subspace. In each iteration, the approximation error will be estimated based on the equation (\ref{FSP_Error}) and the algorithm will not be stopped until the error falls below a previously established threshold. This algorithm is guaranteed to terminate within finite steps if the system is bounded, but it can usually be applied to an infinite system as well. Counterexamples in which FSP algorithm fails do exist (MacNamara, Sidje and Burrage, 2007), though such systems does not usually exist in the physical world. Applications of FSP algorithm for study stochastic noise in gene network can be found in Munsky and Khammash (2008) as well as Neuert \textit{et al.} (2013).

The FSP algorithm can be further augmented in a number of ways, which usually focus on reducing the cost of computing the matrix exponential and optimizing the finite projection space. The Krylov-FSP algorithm (Burrage, \textit{et al.,} 2006) applies an algorithm derived from the Krylov subspace method to improve the computational efficiency. MacNamara, Sidje and Burrage (2007) discussed the importance of the enumerations of state space with respect to the FSP method. Munsky and Khammashb (2007) studied the application of FSP when the initial distribution is sparse, and also proposed splitting the time course so that the support of distribution can be captured more efficiently.  Sunkara and Hegland (2010) discussed the optimality of the projection space's size. Hjartarson, Ruess and Lygeros (2013) proposed a method that combines the stochastic simulation algorithm with the FSP method to reduce the size of finite projection space's size. The FSP method can also be implemented to take advantage of the timescale separation commonly found in genetic systems (Pele\u{s}, Munsky and Khammash, 2006) and can be applied to spatially inhomogeneous stochastic biochemical systems as well (Drawert, \textit{et al.,} 2010). 

The sliding window method (Wolf \textit{et al.,} 2010) also utilizes the idea of reduced state space. The major difference between this method and the FSP method is related to constructing the reduced state space. In particular, given an initial condition, the sliding window method divides the time course $[0,t]$ into multiple segments, and different reduced state spaces named sliding windows are constructed for each segment, so the distribution of then CME system can be calculated sequentially. Compared to the FSP algorithm, the sliding window method requires a smaller reduced state space and thus saves computational resources.
\newline


Another approach to speeding up the estimation of the distribution of a CME system is to aggregate states with similar properties. So, instead of calculating the probability of each state, we only need to calculate the probability of aggregated states, which are fewer in numbers. In fact, the FSP method essentially aggregates the complement of the finite projection space $\boldsymbol{J}$ into a single absorbing state. Such aggregation approaches were firstly developed to study general stochastic dynamic systems (Simon and Ando, 1961; Haviv, 1987; Meyer, 1989). In these early approaches, the state spaces are separated into regions, so that the solution could focus on the interactions between regions rather than on within-region interactions. The aggregation approaches generally work well if the transaction rates matrix exhibits a block structure. 


 A typical aggregation method would utilize two operators: an aggregation operator, which projects the original state space to a smaller aggregated space so that a reduced CME can be constructed and solved, and a disaggregation operator, which can be used to approximate the solution of the original CME using the solution of the reduced CME. QSSA can often serve as a useful guide for setting up these two operators. The aggregation operator can be set up to aggregate different states of the fast species together, and the disaggregation operator can be defined based on the conditional distribution of fast species given the slow species. Applications of such approach can be found in the works of Pele\u{s}, Munsky, and Khammash (2006) and MacNamara, \textit{et al.,} (2008).\newline
  
 \textbf{Example:} Model \ref{Model_Dogma} tracks the copy numbers of both mRNA and protein species, so the state space consists of all non-negative integer pairs $(X_R,X_p)$. The aggregation operator can be defined as $(X_R, X_P)\rightarrow X_R$: all states with the same protein copy number are be aggregated into a single new state. In the aggregated system, the only variable is the copy number of mRNA species, whose equilibrium distribution is Poisson distribution. The equilibrium distribution of the original CME can be then approximated using a disaggregation operator: $\P^s(X_R, X_P)\approx \P^s(X_R)\P^s(X_P|X_R)$ where $\P_s(X_P|X_R)$ represents the quasi-steady-state distribution of protein copy number given mRNA copy number, which is also Poisson distribution. \newline

Other aggregation schemes also exist in the literature. Hegland, \textit{et al.,} (2007) defined an aggregation operator that aggregates every two adjacent states. If such an operator was applied to the same system multiple times, the size of the state space would shrink exponentially. The disaggregation operator was defined to redistribute the probability mass equally over states that are aggregated. This approach approximated the solution of CME with a stepwise distribution and can be applied to a high-dimensional system based on a sparse grid. Waldherr, Wu and Allg\"{o}šwer (2010) applied an aggregation method to study a bistable genetic switch with two activators. To calculate the probability that both genes express at a high level, all states in which the sum of copy numbers of both genes is greater than a certain threshold were aggregated as a single ''on" state. The estimated transition probability into the ``on" state can provide an explanation of how certain long-term characteristics (such as the slow transition into a ``sinking" state) can be generated by reactions that take place on a much shorter timescale. 
\newline

While the state reduction method such as FSP and aggregation methods can reduce the complexity of state spaces and simplify the computation, the discrete nature of the CME can still pose a formidable challenge when the dimension is high. An alternative approach is to apply the hybrid stochastic-deterministic method and approximate part of the CME system with a continuous model. Similar to the QSSA, in a hybrid approach, the system vector $\bold{X}$ is partitioned as $(\bold{Y}, \bold{Z})$ and the solution to the CME is decomposed as  $\P(\bold{X})=\P(\bold{Y})\P(\bold{Z}|\bold{Y})$. Here $\bold{Y}$ represent the species with low copy numbers and are treated as discrete variables. $\bold{Z}$ represent species with high copy numbers and are treated as continuous random variables. $\bold{Y}$ can then be modeled using a reduced CME equation (\ref{QSSA_discrete}) in which variables $\bold{Z}$ are replaced by their conditional expectations. Correspondingly, the evolution functions of $\bold{Z}$, are usually represented using coupled deterministic equations. 

In the first hybrid method for solving the CME (Hellander and L\"{o}tstedt, 2007), $\bold{Z}$ are assumed to be independent random variables following Normal distributions. The variances are assumed to be small, and the expectations are modeled with deterministic differential equations determined by the distribution of $\bold{Y}$. This hybrid system was solved numerically, with the distribution of $\bold{Y}$ approximated by samples generated by the SSA and the expectations of $\bold{Z}$ calculated through a deterministic method. Similar approach was also explored in Henzinger \textit{et al.,} (2010). In Menz, \textit{et al.,} (2012), the explicit evolution functions of the distributions of discrete variables and the expectations of continuous variables are derived using the Laplace's method of integral approximation. 


In the aforementioned hybrid methods, the impact of continuous species $\bold{Z}$ on the evolution of the system is limited to the first moments. Such an assumption may not always be appropriate, especially when the copy numbers of corresponding species are of moderate size or the distributions of the system contain multiple modes. This limitation can be addressed using the method of conditional moments (Hasenauer, \textit{et al., 2014}) in which the evolution functions of the higher moments of continuous variable $\bold{Z}$ conditional on the discrete variables are taken into consideration. This approach essentially combines the key principles behind both the hybrid model and the moment-based approach; thus, it potentially offers a better approximation of the underlying system. 

Another approximation approach worth mentioning is the mean field approximation (Kim, Lepzelter and Wang 2007; Kim and Wang 2007). This approach assumes that the distribution of the system state equals the product of the marginal distribution of individual species and thus avoids the difficulty of handling the interacting term in the CME. In spite of the simplification made in this assumption, certain key traits of the system (such as the probability of the activation of genes and the fluctuation of the molecules) can still be preserved. 
\newline

Despite of the challenge of overcoming the discreteness and the analytical-intractability of the CME system, the various methods discussed in this section provide quantitive tools that can be used to investigate the properties of the CME system and establish relationship between model parameters and the distribution of system state. Still, as most of the methods discussed here are derived based on approximated assumptions, the issue of balancing the computational efficiency and the degree of approximation could be crucial for the proper. In particular, one question that remains to be answered is to what extend the special properties of the CME system can be preserved in a given  approximated method.  

\section{Inference Approach}

We have discussed how to construct a CME-based model, as well as the available methods for studying the properties of the CME. In practice, as direct observations of the inner mechanisms of cellular systems and the values of parameters are rarely possible, we often have to rely on inference approach to understand such complex systems based on observable information. In the case of the gene regulatory systems, the modern experimental technologies may allow scientists to obtain the following information: a) single-cell-level mRNAs or proteins expressions measured at (presumably) steady state; b) snapshots of single-cell-level mRNAs or proteins expressions collected at various time stages, using the samples from the same populations; c) the temporal tracking of mRNAs or proteins molecules of individual cells. From the perspective of the CME model, such observations yield direct information on the equilibrium distribution, the temporal distribution at different times or realizations of the trajectories of the CME. In this section, we will discuss the existing inference approaches that can utilize these information to draw meaningful conclusions about the underlying model. 
\newline

Whether the given model can be solved analytically often serves as an important factor in determining available inference approaches. If the exact analytical solution is available, likelihood-based approaches can be applied to infer the unknown parameters in a CME system.  For instance, the analytical solution of the distribution of the copy number of mRNA molecule in Model \ref{2S_Model} is derived by Raj, \textit{et al.,} (2006). The maximum likelihood approach can then be applied to estimate the key parameters, such as the rates of activation and deactivation of gene (Raj, \textit{et al.,} 2006; Tan and van Oudenaarden, 2010). 

Still, as we will discussed later, when a dynamical system such as the CME is involved in inference problems, it is often necessary to take additional steps to verify whether the estimated model can adequately fit the data and whether the model parameters are identifiable given observed information (Tan and van Oudenaarden, 2010). Raj, \textit{et al.,} (2006) assessed the adequacy of the fitting of Model \ref{2S_Model} by comparing the experimental data and the  simulated samples generated by the estimated model. Zenklusen, Larson and Singer, (2008) suggested that a thorough search of parameters space should be conducted to see if the model can be fitted with different sets of parameter values. \newline

As analytical solutions are rarely obtainable in practice, it is often necessary to introduce additional assumptions to simplify the CME system so that a similar likelihood approach can be applied. One notable example is the distribution of the copy number of protein in Model \ref{Model_Dogma}, which has no analytical solution. However, analytical solution exist regarding the quantitive properties of the mRNA molecule. In particular, the lifetime of an mRNA molecule follows Exponential distribution. As an mRNA molecule produces protein molecules in a constant-rate Poisson process throughout its lifetime, the number of proteins produced by an mRNA molecule follows Geometric distribution. Considering the fact that the lifetime of an mRNA molecule is often much shorter than to the lifetime of a protein, instead of using a Poisson process to model the translation process, we may assume that any mRNA molecule will simply translate a Geometrically distributed number of protein molecules in an instant burst before it degrades. 

Paulsson and Ehrenberg (2000) constructed a CME model based on this principle and showed that, conditioning on the copy number of mRNA, the stationary distribution of the copy number of protein protein follows Negative Binomial distribution. Furthermore, by applying continuous approximation and ignoring the fluctuation in the copy number of mRNA, the distribution of protein copy numbers can be shown to follow Gamma distribution with two parameter $a$ and $b$ (Cai, Friedman and Xie, 2006):
$$p(x)\propto x^{a-1}e^{-x/b}.$$

\noindent Compared to the notations we used in Model \ref{Model_Dogma}, parameter $a$ is roughly equivalent to $\tau_R/\lambda_P$, and can be interpreted as the frequency of burst. $b$ is equivalent to $\tau_P/\lambda_R$ and represents the average number of protein molecules produced per burst. This model can be easily applied to fit single-cell-level protein expression data, and was used to study the expressions of $\beta-$galactosidase in living \textit{Escherichia coli} cells where the ``burst" pattern can be clearly observed (Cai, Friedman and Xie, 2006). Moreover, this model explains the two distinctive patterns of observed distributions of protein expressions: an exponentially-decay-shaped pattern peaked at zero and a bell-shaped pattern with a non-zero peak. A further system-wide examination of the expressions of different protein species in \textit{Escherichia coli} cells (Taniguchi, \textit{et al.,} 2010) also demonstrated that the empirical distributions of different protein species are consistent with the two-parameter Gamma distribution.
 \newline

The above strategies can only be applied to specific models, and a more general method is needed to counter the intractability of distribution functions of the CME systems. One common approach is to focus on the equality or differential equation of the moments of variables of interest. Such moment-based methods may not be able to fully utilize the observed information as we might expect for a likelihood-based approach, but can still provide valuable insight that cannot be obtained by approaches that ignore the intrinsic noise.  For instance, in Model \ref{Model_Dogma}, in order to estimate the parameters and initial conditions, we only need to measure the first two moments of protein and mRNA copy numbers at two separate times if the CME model is used. On the contrary, the number of measurements needed to identify those parameters under an equivalent deterministic system is far greater (Munsky, Trinh and Khammash, 2009).

In relatively simple models such as Model \ref{Model_Dogma} and Model \ref{2S_Model}, it is often possible and sufficient to derive equalities only involving the first two moments for purpose of inference. To estimate the parameters in Model \ref{2S_Model}, So, \textit{et al.,} (2011) derived the analytical formulas of the Fano factor and the square coefficient of variation (variance divided by the square of mean) and fitted them to the observed values. A similar method was used by Gandhi, \textit{et al.,} (2011) to study the coordination between genes during cell divisions, where the dependence between genes introduced by cell division was explored by numerically fitting the analytical formula of the covariances between gene expressions. Such moment-based approaches can also be used to distinguish different model assumptions. In the work of Singh \textit{et al.,} (2012), the Fano factor is used as the major criterion for determining whether the noise at protein level is mainly contributed by the Poisson fluctuation in RNA numbers or by the stochastic transitions between different states of gene. 

In more complicated models where the exact solutions of moments are not available, moment closure methods are usually applied to establish approximated formulas of key moments. Such strategies are especially important for analyzing biological models with rational propensity functions such as Model \ref{Simplified_Rational_Model}. Pedraza and van Oudenaarden (2005) investigated a three genes system in which the interactions are modeled by Hill functions and established moment equalities by taking linear expansion around the steady state. Achimescu and Lipan (2006) and Raffard et al. (2008) studied the inference problem with rational function in a single-gene system with both mRNA and protein species under the presence of external signals. Chao and Wong (in preparation) developed moment-closure inference approaches for analyzing multiple-gene systems and demonstrated that such approach  not only can be applied to estimate the unknown parameters but also can be used to infer the unknown regulatory relationships. In the scenarios where the gene expressions within single cell can be tracked over time, Milner, Gillespie and Wilkinson (2013) proposed to model the observed data as Gaussian distributed random variables whose means and variances are determined by a moment closure scheme. K\"{u}gler (2012) also considered a similar approach but focused on fitting the parameter through minimizing the distance between observed moments and the moments predicted by the estimated model. \newline

%
%
%
%
%

While the moment-based approach provides means to obtain point estimators of the unknown parameters, additional approaches are still needed to quantify the estimation uncertainties.  Zechner, \textit{et al.,} (2012), argued that, due to the large number of cells measured simultaneously in cytometry experiments, it is reasonable to apply the central limit theorem and assume that the empirical moments would follow Normal distributions whose means and variances can be expressed as functions of moments. Consequently, as long as a suitable moment closure scheme can be used to establish quantitative relationships between moments and unknown parameters, the uncertainty of estimations can be quantified using either the frequentist or the Bayesian likelihood-based method. Similar ideas can also be found in the work of Ruess, Milias-Argeitis and Lygeros (2013), Ruess and Lygeros (2015), Fr\"{o}hlich, \textit{et al.,}(2016) and Schilling, \textit{et al.,} (2016).

As we might expect, moment-based approaches that utilize moment-closure schemes would inevitably introduce error to the inference procedure due to the discrepancy between the chosen closure scheme and the true CME system. Unfortunately, such error is often hard to evaluate in practice. Schilling, \textit{et al.,} (2016) proposed an adaptive algorithm to handle this issue. Their approach utilizes a simulation algorithm to generate samples of the estimated CME model, where the parameter values are inferred based on various moment closure schemes. The discrepancy between the simulated samples and observations can then serve as the criterion of choosing the best moment closure schemes. This adaptive algorithm not only can select the most appropriate moment closure schemes for a given model, but also is able to adopt different schemes in different regions of parameter space. \newline

The inference approaches we discussed above all make assumptions about the availability of certain analytical formulas, either exact or approximated, that serve as links between observed data and unknown parameters.  While such approaches are often relatively easy to implement computationally, they do suffer from several potential drawbacks. First, we need to be able to establish such analytical formulas. Second, it is hard to quantify the bias and uncertainty introduced by the discrepancy between approximation formulations and the true model. Third, approximation formulations often focus on certain summary statistics (such as the moments) of the observed data and thus may not be able to fully utilize the information. In the following paragraphs, we will investigate the inference approaches that utilize numerical methods or simulation algorithms to bridge the gap between the data and model parameters.

In principle, the unknown parameters can be estimated by searching the parameter space with the goal of minimizing the discrepancy between the numerical solution of the estimated CME and the observed distribution. The FSP approach discussed in the previous section provides a means to calculate the numerical solution of the CME and can be applied to the both stoichiometric network (such as Models \ref{Model_Dogma}, \ref{2S_Model} and \ref{Repressor_Model}) and the system with the rational propensity functions (such as Models \ref{Nonlinear_Repression} and \ref{Simplified_Rational_Model}). In Munsky, Trinh and Khammash (2009), a CME model similar to Model \ref{Nonlinear_Repression} is used to describe the activity of \textit{lac} operon in \textit{Escherichia coli.}. Parameters in this model were fitted using an optimization algorithm that aimed to minimize the $L_1$ distance between the FSP solution and observed distribution. The parameter values are assigned randomly at the beginning of the optimization procedure and then updated in gradient-based and simulated annealing searches. Similar method can also be found in Neuert, \textit{et al.,} (2013), Shepherd, \textit{et al.,} (2013) and Senecal, \textit{et al.,} (2014). \newline

In the above FSP-based inference method, the numerical solution of the CME must be recalculated throughout the optimization algorithm. Despite of the fact that the FSP method can significantly reduce the time of calculating the numerical solution of the CME, such a task can still be very demanding computationally, even when the dimensions of parameter space are moderate. Then it can be worth considering the likelihood-free inference approach to avoid the need to compute the numerical solution of the CME. 

The Bayesian method known as approximate Bayesian computation (ABC) (Tavar\'eŽ \textit{et al.,} 1997; Pritchard \textit{et al.,} 1999; Beaumont, Zhang, and Balding, 2002) can be used for such a purpose. In a standard ABC rejection algorithm, a particle $\theta^*$ is firstly sampled from the prior distribution of unknown parameters and is used to generate a simulated data set $X_{\theta^*}$. The proximity between $X_{\theta^*}$  and the observed data set $X$ can then be evaluated based on a chosen distance metric. Certain summary statistics might also be used when the dimension of data is high (Jiang \textit{et al.,} 2017). The decision on whether to reject or accept the particle $\theta^*$ will then be made based on whether the distance is greater or smaller than a predefined threshold $\epsilon$. The acceptance rate of particles can be improved by adopting the Markov chain Monte Carlo method (Marjoram \textit{et al.,} 2003 ) and sequential sampling technique (known as ABC SMC, see Toni, \textit{et al.,} 2009, Liepe, \textit{et al.,} 2014). This procedure essentially allows us to obtain independent samples of $\theta$ from density $p(\theta | d(X,\hat{X}) < \epsilon)$, which can be regarded as a reasonable approximation to the posterior distribution $p(\theta| X)$ for small $\epsilon$. 

Thus, as long as we can simulate samples from the given model with specified parameters, we can obtain posterior samples of parameters without evaluating the likelihood function. Considering the numerous stochastic simulation algorithms available for CME systems (section 4), ABC can be a potentially powerful choice for inferring parameters in the CME.


Lillacci and Khammash (2013) developed an ABC algorithm named INSIGHT for analyzing single-cell-level flow cytometry data under a CME-based model. This work explores the issue of the control of false rejection error when applying the ABC algorithm in the context of single-cell-level data. In particular, the decision to reject or accept the proposed particle should be made based on the comparison between the empirical distribution of observed data and the theoretical distribution of proposed model. However, in practice, as the theoretical distribution of the proposed model is often approximated by the simulated samples, false rejection can occur even when the theoretical distribution of the proposed model is consistent to with the observed data. Such error can, in principle, be reduced by increasing the size of simulated data at the price of computational time. Lillacci and Khammash showed that, if the Kolmogorov distance is used as the distance metric, a reasonably low probability of false rejection error can be attained with a relatively small simulated sample size, as long as the size of observations data is large, which is often the case in typical flow cytometry experiments. This discovery shows that the ABC algorithm can be implemented in a very efficient manner for modern single-cell-level experiments. This approach also allows the estimation of a mismatch index, defined as the distance between the empirical distribution of observation and the distribution of best-fitting model. This index grants us valuable insight on the discrepancy between experimental data and the stochastic model, and can be used to determine whether we should investigate alternative models.

As a Bayesian method, ABC opens the possibility of using the Bayes factor or posterior probability to compare competing models (Toni, \textit{et al.,} 2009, Liepe, \textit{et al.,} 2014). For instance, in Toni et al. (2012), the ABC SMC algorithm is used to compare different candidate models that represent different hypotheses of the underlying system based on the posterior probabilities. Moreover, by comparing the simulated samples and the observed data, the ABC method provides means to diagnose the discrepancy between the model and the data (Ratmann, \textit{et al.,} 2009).
\newline

The approaches we discussed above are often designed to study data that represent the empirical distribution of single-cell-level gene expression at steady state or at various stages of evolution. When the expressions of mRNA or protein molecules within individual cells can be tracked continuously (Golding, \textit{et al.,} 2005, Yu, \textit{et al.,} 2006), the corresponding inference problem can be handled in a quite different fashion. In particular, as laid out in the Gillespie algorithm (Gillespie, 1977), as long as we have the complete information on a particular trajectory of the CME over time $[t_0, t_n]$ (including the initial copy numbers $\bold{x}(t_0)$ as well as the firing times of each reactions up to time $t_n$), we can easily express the likelihood function as the product of exponential and multinomial densities.The corresponding inference problem can then be easily solved. For instance, in a stoichiometric system where the propensity functions are linear functions of the unknown parameters, the maximum likelihood estimator can be solved analytical given the full trajectory (Daigle \textit{et al.,} 2012).

Nonetheless, the complete information is extremely difficult, if not impossible, to obtain. In practice, we can only observe the system state at a few discrete time points. Let us represent the observed data as $\bold{x}(t_0), \bold{x}(t_1),\cdots, \bold{x}(t_n)$. The likelihood function is then the products of transition likelihood $p(\bold{x}(t_i) |\bold{x}(t_{i-1}), \theta)$ whose expression is usually not analytical.  For example, the mRNA in Model \ref{Model_Dogma} evolves as a simple birth and death process. Suppose the copy numbers are 10 and 20 at time s$0$ and $t$, respectively, then any full trajectory that satisfies the following conditions is consistent with the observed data: 1) the total number of births minuses the total number of deaths during $(0,t]$ equals 10; 2) the birth events and the death events can occur in any order and at any time as long as the total copy number never drops below 0. Consequently, transition probability from $0$ and $t$ will be the sum of probabilities of all the consistent full trajectories, which can be hard to compute if the system is complex. 

Many authors have thus explored approximated approaches to estimate the transition probability so that the unknown parameters can be inferred with conventional methods. Reinker, Altman and Timmer (2006) discussed a strategy for estimating the transition probability in stoichiometric networks. Assuming that the number of firings is limited or the propensity functions remain constant during period $(t_{i-1}, t_{i}]$, the transition probability can be approximated with relatively simple analytical formulas. This approach is roughly equivalent to approximating the exact transition probability by excluding the less probable paths. Tian \textit{et al.} (2007) estimated the transition likelihood from $t_{i-1}$ to $t_{i}$ based on simulated realizations of system $t_{i}$ through non-parametric kernel density function.

More generally, the maximum likelihood estimator of parameters can be found using the expectation-naximization (EM) algorithm if we treat the full trajectory as complete data. In the E-step, the expectation of the likelihood function of the full trajectory, given the observations and current value of parameters, is evaluated. In the M-step, parameter values are updated by maximizing the conditional expectation. As it is usually impossible to calculate the exact conditional expectation in the context of a CME system, the Monte Carlo extension of the EM algorithm (MCEM; Wei and Tanner, 1990) is often used in practice. In MCEM, the conditional expectation is estimated based on the sampled full trajectories. The major difficulty in applying MCEM in a CME system lies in the fact that the simulated trajectories must be consistent with the observed data, which can be hard to achieve if we use an unmodified stochastic simulation algorithm. Horv\'ath and Manini (2008) suggested that the full path should be simulated piece-wisely for each interval $(t_{i-1}, t_{i}]$.  Daigle \textit{et al.}, (2012) argued that in order to implement MCEM efficiently, the initial choice of parameters should be the values that are likely to generate consistent trajectories.  An iterative algorithm based on the cross-entropy method (Rubinstein, 1997) was used to find such initial values. In each iteration, trajectories are simulated using previous parameter values but only the trajectories that are closed to the observed path are used for updating the parameters. 

Wang \textit{et al.} (2010) discussed an approach in which likelihood function is maximized using stochastic gradient descent. It was shown that the gradient of the likelihood function can be determined based on the expectation of the system's durations in different states and the numbers of transitions between states, conditioning on the observed path.  A reversible jump Markov chain Monte Carlo algorithm is implemented for simulating paths that are consistent with the observations, in which new paths are proposed by adding or deleting certain sets of reactions from the initially proposed path. This method can be applied to the data set with only some of the species observed.

In addition to the frequentist approaches, Bayesian methods that utilize MCMC sampling algorithms can also be used to solve such problems.  Boys, Wilkinson and Kirkwood (2008) use the MCMC algorithm to sample the full trajectories conditioning on the observations. The efficiency of MCMC sampling is improved by using reversible jumping and blocking update methods. Generally speaking, the Bayesian approach can usually be directly applied to a system with unobserved species, as the Bayesian approach can simply impute such missing information in the same way as imputing the full trajectories. \newline

As we have seen in the above discussion, the discreteness of the CME is often the major obstacle for estimating the transition probability. Another way to handle this issue is to use continuous approximation, including LNA and SDE. In the following paragraphs, we will discuss existing inference approaches utilizing the continuous stochastic models. 

LNA approximated the CME as the sum of the deterministic term and stochastic fluctuation. As noted in Komorowski \textit{et al.} (2009), the stochastic fluctuation can be modeled by SDE whose drifting and diffusion terms depend on the deterministic part of LNA. Consequently, the solution of LNA is always multivariate Gaussian distribution with both mean vector and covariance matrix determined by the propensity functions. Thus, with suitable choice of prior over the unknown parameters, the posterior distribution can be sampled straightforwardly using the standard Metropolis-Hastings algorithm. This framework can readily accompany the presence of unobserved species, as well as the measurement errors (such as additive Gaussian noise). This method was applied to estimate the GFP protein degradation rate. Fearnhead, Giagos and Sherlock (2014) also considered the inference problem using LNA and showed that such an approach can be statistically and computationally more efficient than approaches based on deterministic differential equations. 

Unlike LNA, the transition probability between two successive observations in a full SDE approximation is often analytically intractable. However, such transition probability can often be estimated by discretizing the trajectory of an SDE system, a scheme commonly known as Euler-Maruyama approximation. This approximation discretizes the sample path between two successive observations into multiple segments, and the increments in each segment are modeled as independent Gaussian random variables whose means and variances are determined by SDE. This approximation forms the basis of the Bayesian inference framework proposed by Golightly and Wilkinson (2005) for general stoichiometric models. An MCMC scheme is then used to obtain posterior samples of unknown parameters. Due to the need to impute values to discretize the SDE and handle the unobserved species, the sampling procedure alternates between the sampling of parameters conditional on the augmented data and the sampling of missing data given observations and the current set of parameters. 

This scheme can be further enhanced with advanced sampling methods. To overcome the dependence between the parameters and missing data, sequential MCMC methods can be used to sample the model parameters (Golightly and Wilkinson, 2006). The accuracy of Euler-Maruyama approximation increases as the number of imputed values increases. However, if we increase the number of imputed values, we will also increase the computational cost. Golightly and Wilkinson (2008) proposed a global MCMC strategy with an improved Gibbs sampler, in which a Brownian motion process is used to impute values between successive observations so that the computational cost will not scaled up as the number of segments increases. The speed of such a strategy is still limited by the complexity of the model, and the particle Markov chain Monte Carlo method can be implemented to lessen the computation burden (Golightly and Wilkinson, 2011)
\newline

As we have discussed in section 2, it is possible to propose a model with different levels of detail to explain the mechanics of gene regulatory systems. Nonetheless, it is often hard to observe the inner mechanism of a regulatory system directly, and we have to rely on the available information to choose between different models. For instance, how shall we choose between the simple Model \ref{Model_Dogma} and the two-state Model \ref{2S_Model} by analyzing the single-cell-level distribution of the copy number of protein molecules? And how can we know if our choice of model is sophisticated enough to explain what we have observed or whether the observed information is sufficient for us to infer the details of the model we propose? In the following paragraphs, we will review the relevant literatures that deals with the problem of model selection in the context of studying CME systems as well as general complex dynamic systems. 

To evaluate the fitness of a model, suitable metrics are needed to measure the distance between model prediction and observed data. In the literature, the distances between the observed and predicted values of time derivatives of state variables or between the predictive and observed moments are often used (K\"{u}gler, 2012; Babtie, \textit{et al.,} 2014; Liepe, \textit{et al.,} 2014). If the predictive distribution can be obtained, $\chi^2$ test can be used to determine whether the prediction of a model is consistent with the data (Zenklusen, Larson and Singer, 2008), and Euclidean distance or Hellinger distance can be used as a metric of discrepancy (Munsky, Trinh, and Khammash, 2009; Sunn\aa ker, \textit{et al.,} 2013; Silk,  \textit{et al.,}  2014). To compare models with different level of detail, AIC as well as the Bayes factor can be used to penalize additional model parameters or structures (Toni, \textit{et al.,} 2009; Sunn\aa ker, \textit{et al.,} 2013; Babtie, \textit{et al.,}2014; Liepe, \textit{et al.,} 2014; Silk, \textit{et al.,} 2014). The fitness and complexity of the model can also be balanced by measuring the uncertainty introduced by an over-complex model. For instance, in the work of Neuert,\textit{et al.,} (2013), the log-likelihoods are used as a measurement of fitness while uncertainty is evaluated by cross-validation. Specifically, the uncertainty is defined as the average log-likelihoods of a complete data set calculated using parameters estimated based on sampled partial data sets. Then the best model is chosen by balancing both fitness and uncertainty. 

As many authors have pointed out, due to the complexity of dynamic system, even if we can achieve a good fit using a particular model or particular set of parameters, there might be other alternative models or sets of parameters that could fit the data equally well. Consequently, it is often useful to search the parameter space or even the space of candidate models thoroughly before making a final conclusion. Villaverde, \textit{et al.,} (2015) explored the predictive accuracy of the fitted model using a consensus approach. This approach searches the parameter space and records sets of parameter values that all fit the data well. Then the accuracy of prediction can be analyzed based on whether these collected sets of parameter values could reach consensus or not. The burden of searching the parameter space can be reduced by grouping the parameters into modules of meta-parameters. Uncertainty in model structure is explored in the so called topological sensitive analysis (Babtie,  \textit{et al.,}  2014). In this approach, alternative structures are proposed by modifying the relationship between nodes in the current model. Restrictions are imposed to limit the search space. The fitness of the proposed structure is then evaluated using Gaussian process regression. The topological filtering method (Sunn\aa ker, \textit{et al.,} 2013) explores alternative models by constructing a tree of models. The root of this tree is the base model and consists of the most detailed interactions. New nodes along the tree are then created by removing interactions and the associated parameters gradually. Analysis of the fitness of the new model is conducted at the same time. And the process of creating new nodes is stopped if any further simplification will make the model unfitted for the data. This approach could lead to a tree with multiple branches and the candidate models at the top of each branch are collected for in-depth study.  Finally, the exploration of alternative models may guide the scientists to design new experiments to further discriminate among different models. Maximizing Fisher information has been proposed as the metric guiding the design of new experiments for understanding biochemical reaction network (Ruess and Lygeros, 2013). The interaction between inference, model selection and design of experiment also served as the central theme in the 2014 DREAM contest (Meyer \textit{et al.,} 2014). \newline

\section{Discussion}

In this review article, we explored the CME-based approach in modeling and analyzing the intrinsic noise in gene regulation systems. We demonstrated that the CME-based model offers a flexible way of incorporating physical mechanisms and is capable of capturing the discrete, stochastic and dynamical nature of cellular systems. We also discussed several alternative modeling approaches and viewed them as approximations to the equivalent CME model. We then provided an overview of available tools that can be used to study CME-based systems. In particular, we discussed simulation methods, analytical and numerical solvers and statistical methods that can be used to infer the unknown parameters or structures in CME models based on data collected at single-cell level. 

Our discussion in this article has been focused on the intrinsic noise. It must be admitted, though, the systems we considered are also under the influence of extrinsic noises. Alll the single-cell-level experiments are subjected to measurement errors. In the smFISH approach, as individual molecules can be visualized and tracked, it is possible to obtain the discrete counts of the molecular species under investigation. However, in technologies such as flow cytometry and mass cytometry, single-cell-level observations are collected by measuring the intensity signals emitted by reported tags attached to target molecules. Consequently, we would need to infer discrete counts of molecules from the observed continuous intensity signals so that CME model cann be applied. In practice, it is often assumed that the observed intensities follow Gaussian distributions whose means and variances are proportional to the counts (Munsky, Trinh and Khammash 2009; Komorowski \textit{et al.}, 2009; Lillacci and Khammash, 2013). Other models exist as well, such as Gaussian random variables with constant variances (Golightly and Wilkinson, 2011). In scRNA-seq experiments, how to properly normalizing the observed read counts and correcting bias introduced by the unobserved and dropout measurements also pose serious challenges for data analysts (Stegle, Teichmann, and Marioni, 2015; Bacher and Kendziorski, 2016). 

In addition to measurement noises, systems at single-cell levels are subjected to various fluctuations induced by external factors. With carefully designed experiments, it is possible to distinguish intrinsic and extrinsic noise. For instance, in the study by Elowitz, \textit{et al.,} (2002), single-cell-level expressions of two different protein molecules driven by the same promoter were observed. The extrinsic noise is defined as the fluctuations that impact the expressions of both proteins simultaneously. Still, for general cases, we need to explicitly model the external factors to distinguish between different sources of noises.  For instance, the fluctuation introduced by cell division can be modeled based on the assumption that molecules in parent cells are randomly distributed to offspring cells upon division (Rigney, 1979; Rosenfeld, \textit{et al.,} 2005). Cell growth, on the other hand, increases the cell volumes and thus decreases the concentrations of molecules. Then we may model the fluctuations introduced by cell growth in a way similarly to the modeling of intrinsic noises introduced by the degradation of molecular species (Lei, \textit{et al.,} 2015). Moreover, cells in the same population are often originated from common ancestors and share information through extracellular communication. A a full understanding of the regulatory interactions at single-cell level may only be achieved by taking the population context into consideration (Snijder and Pelkmans, 2011). 

Suffice to say, the story told by intrinsic noise is not complete. Nonetheless, information extracted from intrinsic noise can still allow us to understand the functions of the fundamental building blocks of cellular systems. In this way, the work on the intrinsic noise will provide a foundation stone for constructing constructing more sophisticated models and serve as an integral part of the quest to understand life.

This review is restricted to CME based approaches for analyzing intrinsic noise in gene regulation systems. While such approaches does have the advantage of providing a detailed, physically interpretable model, they also presents greater challenges for both modeling and inference. On the one hand, many of the approaches discussed in this article have been applied to study relatively small and compact systems with various degrees of success. On the other hands, how such approaches can be applied to understand larger systems involving hundreds and thousands of molecular species and biochemical reactions is still an interesting question without a definite answer. Still, we would like to point out that the knowledge gained by analyzing small systems via CME-based approaches can often provide valuable insight in understanding more complex systems.  

For instance, under the simplest Model 1 we can show that the Fano factor of the copy number of mRNA should equals 1 and the Fano factor of the copy number of protein is greater than 1. This simple conclusion suggests that the Fano factor can be used to measure the strength of intrinsic noise and any deviation from baseline indicates more complicated mechanisms (Thattai and van Oudenaarden, 2001; Tao, 2004).  The study on the Fano factor of molecule species in different cells has reveal many fundamental characteristics of cellular systems. For instance, it has been observed that, in prokaryotic cells, the fluctuation of protein is often determined by and positively correlated with the translational efficiency (Ozbudak, et al. 2002). And in eukaryotic cells, strong correlation between noise and transcriptional efficiency can be found (Blake et al., 2003). In addition, the regulatory pathway would also impact the strength of intrinsic fluctuations, and it is known that the negative feedback loop can reduce the noise and positive feedback can increase the noise (Kepler and Elston, 200; Becskei and Serrano, 2000; Isaacs et al., 2003). 

Taking this discussion one step further, the study on two-gene systems suggests that the correlation between the expressions of different genes measured at the single-cell level can be used as an indicator of an underlying regulatory relationship. Simple analysis would show that correlation between genes tends to be negative if one gene represses the expressions of another gene, and positive in case of activation. In the traditional ensemble based experiments, similar information can only be obtained through the introduction of perturbation. In the work of Stewart-Ornstein, Weissman and El-Samad (2012), such a principle is used to categorize 182 studied proteins in yeast cells based on the correlation matrix measured at steady-state. It was observed that the genes within the same block often have similar functions and respond to the same upstream regulators. This study also found evidence of the correlation between intrinsic noises and external stimulus, which suggests that the observed intrinsic fluctuation can be used to study the regulation pathway. In addition, if the observations can be made at different time points, then the dynamical cross-correlation function can be used to determine the direction of regulatory relationships, as the change in the expression of the upstream gene will only impact the expression of the downstream genes after delay (Dunlop et al, 2008). 

To conclude, the CME-based approach provides a unique and indispensable perspective in understanding the role of intrinsic noises in cellular systems. More importantly, the discrete and dynamical natures of the CME also present fresh challenges for statisticians. First, as we have discussed in this article, the intrinsic noises contains valuable information that can only be extracted via a physically interpretable model such as the CME. Nonetheless, the introduction of such a model also obscures the relationship between the parameters to be inferred and the observed data. Consequently, it is often necessary to find suitable inference approaches to avoid the evaluation of distribution functions. Then what principle shall we follow so that the proposed method can fully utilize information contained in the data? Second, even for a relatively simple genetic toggle switch model, the stationary distribution of the corresponding CME can be unimodal or bimodal. For a CME with bimodal stationary distribution, its evolution through time would exhibit phase transitions between unimodal and bimodal distributions. The low-copy-number effect of cellular systems also forces us to consider discrete distributions that can not simply be approximated as continuous distribution. How well do the traditional inference approaches fare with such unconventional distributions? Third, in the context of complex dynamical models such as the CME, there are no satisfactory answers on how to determine whether the unknown parameters can be identified solely based on available information, or how to compare competing models with different levels of detail. Can we adjust conventional model selection approaches or do we need to develop fresh new methods to solve such problems? In this regard, we hope that our review article not only introduces to the statistical community existing works about applying the CME-based model to analyze cellular systems, but also ignites new research interests toward this direction. 

\section{References}
\nocite{*}
\printbibliography 

\end{document}